\documentclass[12pt]{scrartcl}

\usepackage[parfill]{parskip}
\usepackage{setspace}

\usepackage[natbibapa]{apacite}
\usepackage{amssymb,amsmath}

\usepackage{graphicx}

\usepackage[section]{placeins}
	\makeatletter
	\AtBeginDocument{%
	  \expandafter\renewcommand\expandafter\subsection\expandafter{%
	    \expandafter\@fb@secFB\subsection
	  }%
	}
	\makeatother	

\usepackage{units}
\graphicspath{{./figures/}}
\DeclareGraphicsExtensions{.eps}

\addtokomafont{caption}{\small}
\addtokomafont{captionlabel}{\small}
\addtokomafont{captionlabel}{\bf}
\setcapindent{0em}

\newcommand{\abs}[1]{\left|#1\right|}

\newcommand{\slfrac}[2]{\left.\raisebox{3pt}{$#1$} \! \middle/ \! \raisebox{-3pt}{$#2$}\right.}
\mathchardef\mhyphen="2D

\DeclareMathOperator{\Cin}{Cin}
\DeclareMathOperator{\sinc}{sinc}

\title{Convolution Metric for Neuron Membrane Potential Recordings}
\author{Garrett N. Evans\footnote{gne101 @ psu.edu}\
        \\ { \normalsize \emph{104 Davey Lab, Penn State University, University Park, PA 16802, USA} } }

\date{September 7, 2014}

\begin{document}
\maketitle

\begin{abstract}
I provide a \emph{convolution metric} which takes neural membrane potential recordings as arguments and compares their subthreshold features along with the timing and number of spikes within them---summarizing differences in these with a single ``distance'' between the recordings. Based on \citeauthor{vRossum}'s \citeyearpar{vRossum} metric for spike trains, the metric relies on a convolution operation that it performs on the input data. The kernel used for the convolution is carefully chosen such that it produces a desirable frequency space response and, unlike van Rossum's kernel, causes the metric to be first order both in differences between nearby spike times and in differences between same-time membrane potential values: an important trait.
\end{abstract}

\section{Introduction}
 \label{Ch:dGlobal}

Electrophysiological methods, voltage-sensitive fluorescence \citep[see][]{ScanzianiHausser2009} and computer simulations \citep[e.g.,][]{HellgrenEtAl1992} are all ways that neuroscientists access membrane potential time courses for (real or in silico) neurons and ensembles thereof. In the effort to analyze these signals, a variety of approaches are often taken, including spectral analysis, point-process intensity function estimation, and the peri-stimulus time histogram among others \citep[see][]{MitraBokil2007}. While the results of such analyses provide avenues for signal comparison, it is also convenient to have a measure that compares signals more directly: without recourse to time-binning, a statistical model or other major processing \citep{PaivaEtAl2009}. It is therefore worthwhile to have a metric which, given two membrane potential trajectories, computes a meaningful, non-negative ``distance'' between them. Adding distances between trajectories for individual neurons gives an ensemble distance.

In the case of spike trains, considerable work has been done to develop this kind of measure \citep[see][]{PaivaEtAl2010}. The conventional approach to comparing spike trains has been to use a time-binning procedure to transform trains into finite-dimensional vectors and to evaluate, e.g., a Euclidean distance on the vectors. Time-binning, however, has major drawbacks if precise spike timing is of interest, as several authors note \citep{VictorPurpura97, vRossum, SchreiberEtAl2003}. E.g., it is insensitive to timing changes that do not change a spike's bin, and a timing difference of a single bin is treated the same as a difference of any (nonzero) number. Consequentially, there has been an effort to develop \emph{binless spike train measures}. Two well-known binless measures are the Victor-Purpura metric and the van Rossum metric \citep{VictorPurpura97, vRossum}.

Spikes and relative spike timing are very meaningful components of the membrane potential time course; however, analyses attending exclusively to spikes provide only part of the information available in the signal. As \citet{Lennie2003} has pointed out, energy considerations indicate that neurons spend the bulk of their time in an inactive, non-firing state, i.e., they spend more time ``listening'' than they do ``talking.'' \citeauthor{Lennie2003} estimates about 0.16 spikes per neuron per second in the awake human brain, and suggests that optimally no more than 4\% of neurons are actively firing at any time---which corresponds to a 4 Hz average spike rate for active neurons.

A metric that looks only at spikes ignores information concerning what is happening during inactivity and between spikes during active firing. Such a tool offers a view into what the neuron is ``saying'' but not into what it is ``hearing.'' The latter is certainly of interest to neuroscientists. For example, \citet{SteriadeEtAl1993, AliEtAl1998, BrunoSakmann2006, RosenMooney2006, LongEtAl2010} and \citet{ZiburkusEtAl2006} are just a few studies that look closely at non-spiking features of neural recordings.

\begin{figure}
\centering
\includegraphics{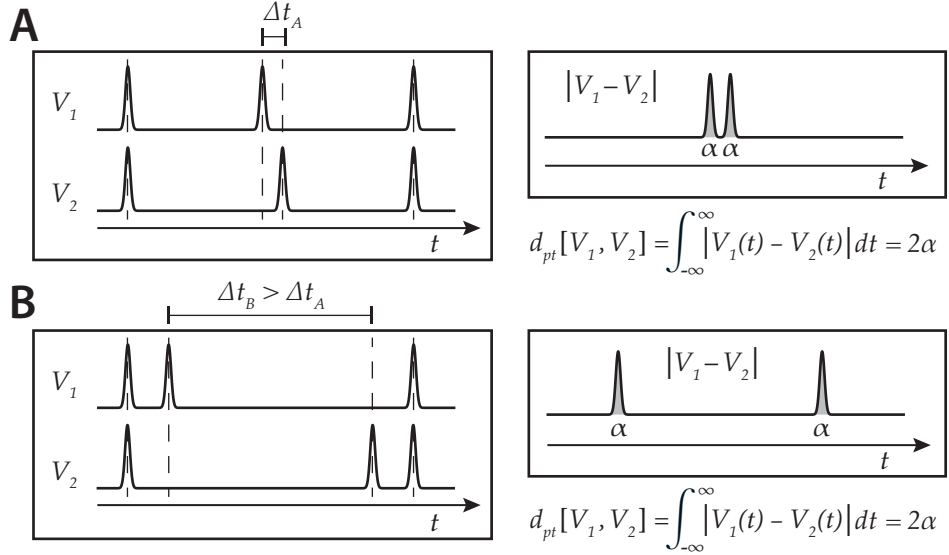}
\caption{{\bf Point-comparison metrics and membrane potentials.} In \emph{A}, the action potential timing is a closer match than in \emph{B}, but according to $d_{pt}$, since the mismatched action potentials do not overlap in either case, both pairs of time courses are the same distance apart.}
\label{Fig:ProblemDPt}
\end{figure}

On the other hand, applying a point-comparison metric to membrane potentials, e.g., $d_{pt}\left[V_1,V_2\right] \equiv \int_{-\infty}^\infty \abs{V_1 \! \left(t\right) - V_2 \! \left(t\right)} dt$, doesn't yield a meaningful comparison of spike timing. This is because non-overlapping action potentials, which are quite ($\sim$1--5 ms) narrow, are treated as equally different by a point-comparison metric no matter how much time separates them. See Fig.~\ref{Fig:ProblemDPt} for an illustration.

We need an integrated membrane potential time course metric that is sensitive both to relative action potential timing and to subthreshold membrane potential dynamics. In this article, I provide such a metric by building off \citeauthor{vRossum}'s metric for spike series \citep{vRossum}. We will generalize the van Rossum metric in such a way that it applies to membrane potential recordings and adapt it so that it has a first-order response to both spike time differences and membrane potential differences. We will also tailor the metric so that it has a fat-tailed, low-pass response to input frequencies that is free of zeros or local minima. The metric is defined in Eq.~(\ref{Def:dC}).

\section{Background}
\label{Sec:Bg}

In the previous section, we discussed two popular binless spike-time sensitive metrics for spike trains: the Victor-Purpura metric \citep{VictorPurpura97} and the van Rossum metric  \citep{vRossum}. We will now briefly review these two results, both of which will figure in what follows.

The Victor-Purpura metric works by assigning costs to basic transformations on spike trains, defining the distance between trains as the cost of the least costly sequence of transformations mapping one train to the other. Three basic transformations are considered: spike insertion, spike removal and spike displacement in time. Spike insertion and removal are both given the same set cost, and spike displacement has a cost that is proportional to the amount of the displacement. This results in a metric, $D_{V\! P}$, that rises proportionally to spike time differences for nearby spikes\footnote{We will use a capital $D$ for spike train metrics and a $d$ for distances on functions of continuous time.}. The distance reaches a hard plateau when it becomes less costly to remove a displaced spike and re-insert it at its new location than it is to move it.

\citet{VictorPurpura97} provide an algorithm that computes their metric with $O(n_1 \cdot n_2)$ computational complexity, where $n_1$ and $n_2$ are the numbers of spikes in the trains. Let $s_1$ and $s_2$ be two spike trains, each consisting of a sequence of spike times:
\begin{align}
\label{Def:SpTrain}
s_1 = \left(t_{11}, \ t_{12}, \ \ldots \ t_{1i}, \ \ldots \ t_{1n_1} \right) ;
\ s_2 = \left(t_{21}, \ \ldots \ t_{2i}, \ \ldots \ t_{2n_2} \right)
\end{align}
\citeauthor{VictorPurpura97}'s algorithm inductively builds up a minimum transformation cost matrix, $\mathbf{G}$, between the trains, the $i j$th entry of which, $g_{ij}$, is defined as the minimum cost of a transformation from the first $i$ spikes of $s_1$ to the first $j$ spikes of $s_2$. With the boundary conditions, $g_{i0} = c_{in} i$, $g_{0j} = c_{in} j$, where $c_{in}$ is the spike insertion/removal cost (taken to be the same cost), the matrix can be built up as:
\begin{equation}
g_{ij} = \min \left\{ g_{\left(i-1\right) j} + c_{in}, \  g_{i \left(j-1\right)} + c_{in}, \ 
	g_{\left(i-1\right) \left(j-1\right)} + q \! \cdot \! \abs{ t_{2i} - t_{1j}}\right\}
\end{equation}
where the constant, $q$, is the proportionality factor for spike-displacement costs. The bottom right entry of $\mathbf{G}$, $g_{n_1 n_2}$, yields $D_{V\!P} \! \left(s_1,s_2\right)_q$: the least cost of a transformation between $s_1$ and $s_2$. \citeauthor{VictorPurpura97}'s algorithm is an adaptation of one \citet{Sellers1974} introduced in the context of a metric for DNA sequences.

The van Rossum distance\footnote{This metric also appears in an earlier publication by \citet{HunterEtAl1998}.} solves the problem another way: First, spike trains are mapped to real-valued functions of time by replacing each spike in a train with a one-sided exponential decay, $H \! \left(t - t_s\right) e^{-\left(t - t_s\right)/\tau}$. Here $t_s$ is the time of the spike being replaced, $H \! \left(x\right)$ is the Heaviside step function, and $\tau$ is a parameter describing the timescale of the metric's sensitivity to spike time differences. The standard $L^2$ norm, $d_{L^2}\left[f_1,f_2\right] \equiv \left(\int_{-\infty}^\infty \left(f_1 \! \left(x\right) - f_2 \! \left(x\right)\right)^2 dx\right)^{\nicefrac12}$, then defines a distance between these functions which, after scaling by a factor of $\nicefrac1{\tau}$, is taken to be the distance between the spike trains. Because the one-sided exponential profile spreads spikes out by the amount, $\tau$, the $L^2$ norm becomes responsive to time differences of order $\tau$ between spikes.

The resulting metric, $D^2_{vR}$, may be stated mathematically in the following way:
\begin{align}
\label{Def:D2vR}
& D^2_{vR}\left(s_1,s_2\right)_{\tau}
\equiv  \frac1{\tau} \int_{-\infty}^{\infty} 
\left( \sum_{i=1}^{n_1} \sigma_{vR}\left(\left(t - t_{1i}\right)/\tau\right) -
\sum_{i=1}^{n_2} \sigma_{vR}\left(\left(t - t_{2i}\right)/\tau\right) \right)^{\! \! 2} \! dt \\
& \mbox{where} \ \sigma_{vR}\left(t/\tau\right) \equiv
	\begin{cases}
		e^{- t/\tau} &:t \ge 0 \\
		0 \ &: \ \mbox{otherwise}
	\end{cases}
\end{align}

The distance squared, $D^2_{vR}$, is preferred to its square root ($D_{vR}$) because, for spike trains that only contain a single spike, we have \citep[pg. 755, Eq.~2.8]{vRossum}:
\begin{align}
\label{d2vR_ss}
D^2_{vR}\left(s_1,s_2\right)_{\tau} = 1 - e^{-\abs{t_d}/\tau}
\end{align}
where $t_d$ is the time difference between the spikes in the two trains. The distance squared increases proportionally with the time difference and levels off as the difference becomes large compared to $\tau$. This means that, for spikes that are nearby compared to the adjustable parameter, $\tau$, $D^2_{vR}$ gives proportional reporting for an interesting piece of quantatative information, the timing difference between the spikes. ($D_{vR}$ reports the square root of the spike timing difference.) The $\tau$ parameter sets the timescale at which timing differences become sufficiently large that further increases are uninteresting.

As reported by \citet[pg.~408, Eq.~6]{PaivaEtAl2010}, in the case of arbitrarily many spikes, the van Rossum metric generalizes to\footnote{Here we are subtracting the third double sum from the overall expression rather than adding it as originally appeared in \citet{PaivaEtAl2010}, which we take to be a mistype.}:
\begin{align}
\label{dvR2_PaivaEtAl}
& D_{vR}^2\left(s_1,s_2\right)_\tau  = \tfrac12 \sum_{i=1}^{n_1} \sum_{j=1}^{n_1}
	e^{-\abs{t_{1i} - t_{1j}}/\tau} + \tfrac12 \sum_{i=1}^{n_2} \sum_{j=1}^{n_2}
	e^{-\abs{t_{2i} - t_{2j}}/\tau}
 	- \sum_{i=1}^{n_1} \sum_{j=1}^{n_2} e^{-\abs{t_{1i} - t_{2j}}/\tau}
\end{align}
This expression allows the van Rossum distance to be calculated with $O \! \left(n_1 \! \cdot \! n_2\right)$ computational complexity. In the third double sum, we see terms that mirror the spike-time comparison term in the single-spike expression (Eq.~(\ref{d2vR_ss})); they report the spike time difference for all pairs of spikes not belonging to the same train. We also see terms that decrease with the timing difference between spikes in the same train. These terms function similarly to the `1' in Eq.~(\ref{d2vR_ss}) in that they exactly cancel the inter-train terms if the trains are identical---so that $D^2_{vR}=0$ in this case. As we will discuss in Sec.~\ref{Sec:MnySpkRec}, they also act to cancel out inter-train spike time comparisons that are the least relevant.

Eq.~(\ref{dvR2_PaivaEtAl}) may be rewritten in two important ways:
\begin{align}
\notag D_{vR}^2\left(s_1,s_2\right)_\tau & = \nicefrac12 \left(n_1 + n_2\right)
	 + \sum_{i=1}^{n_1} \sum_{j=1}^{i-1} e^{-\abs{t_{1i}-t_{1j}}/\tau} \\
	 & \qquad + \sum_{i=1}^{n_2} \sum_{j=1}^{i-1}  e^{-\abs{t_{2i}-t_{2j}}/\tau} 
	 - \sum_{i=1}^{n_1} \sum_{j=1}^{n_2} e^{-\abs{t_{1i} - t_{2j}}/\tau}
	 \label{Eq:D2vRFar} \\
	\mbox{and } D_{vR}^2\left(s_1,s_2\right)_\tau \notag & = \nicefrac12 \left(n_1 -n_2\right)^2
	- \sum_{i=1}^{n_1} \sum_{j=1}^{i-1} 
		\gamma^{\, 2}_{vR}\left(\left(t_{1i} - t_{1j}\right)/\tau\right)^2 \\
	& \qquad - \sum_{i=1}^{n_2} \sum_{j=1}^{i-1}
		\gamma^{\, 2}_{vR}\left(\left(t_{2i} - t_{2j}\right)/\tau\right)^2
	+ \sum_{i=1}^{n_1} \sum_{j=1}^{n_2}
		 \gamma^{\, 2}_{vR}\left(\left(t_{1i} - t_{2j}\right)/\tau\right)^2 \label{Eq:D2vRClose}
\end{align}
where $\gamma^{\, 2}_{vR}\left(t/\tau\right) \equiv 1 - e^{-\abs{t}/\tau}$.

Van Rossum \citeyearpar{vRossum} explored two important limits of his metric, which follow, respectively, from these two expressions. Eq.~(\ref{Eq:D2vRFar}) shows that when all spikes are separated by much more than $\tau$, $D^2_{vR}$ essentially counts spikes in the two series, returning $\nicefrac12\left(n_1 + n_2\right)$. Alternatively, in Eq.~(\ref{Eq:D2vRClose}), we see that if all spike pairs are close together compared to $\tau$, the metric squares the difference in the number of spikes, yielding $\nicefrac12 \left(n_1 - n_2\right)^2$.

Both the van Rossum metric,  $D^2_{vR}$, and the Victor-Purpura metric, $D_{V \! P}$,  share the useful property of rising linearly with spike time differences for nearby spikes and leveling off as spike time differences get large and spikes are (seemingly) unlikely to correspond. They also share the same $O(n_1 \! \cdot \! n_2)$ order of computational complexity.

\section{Methods}
\label{Sec:Methods}

While the Victor-Purpura metric is without question a fascinating and effective tool, it seems difficult in principle to convert its spike-train-transformation approach into an integrated method applicable to membrane potential recordings. Because action potentials do not have hard boundaries and never share exactly the same shape, it is problematic to transform between membrane potentials by cutting and pasting spikes.

By contrast, van Rossum's metric generalizes easily to such a context. As pointed out by \citet{PaivaEtAl2010}, the replacement of spikes with a one-sided exponential decay so central to van Rossum's method is equivalent to convolving the decay with a spike train function consisting of Dirac delta functions serving in the place of spikes:
\begin{align}
&\widehat{f}_s^{\, vR}\left(t;\tau\right) \equiv \int_{-\infty}^{\infty} \sigma_{vR}\left(\left(t-t'\right)/\tau \right) \cdot f_s \! \left(t'\right) dt'  \\
&\mbox{where} \ f_s \! \left(t\right) = \sum_{t_i \in s} \delta \! \left(t - t_i\right) 
\label{Def:fs}
\end{align}
Here, $s$ is a spike train (see Eq.~(\ref{Def:SpTrain})). In these terms, we have:
\begin{equation}
D^2_{vR}\left(s_1,s_2\right)_{\tau} = \frac1{\tau} \int_{-\infty}^{\infty}
	\left(\widehat{f}_{s_1}^{\, vR} \! \left(t;\tau\right)
	- \widehat{f}_{s_2}^{\, vR} \! \left(t,\tau\right) \right)^2 dt
\end{equation}

This way of writing $D^2_{vR}$ suggests an easy adaptation to membrane potentials. Where $V \! \left(t\right)$ is the membrane potential of a neuron as a function of time and $T$ is the (finite) time domain over which the neuron has been recorded from, one may define:
\begin{align}
\widehat{V}^{\sigma_{vR}} \left(t;\tau\right)_T = \frac1{\tau} \int_T \sigma_{vR}\left(\left(t-t'\right)/\tau \right) 
V \! \left(t'\right) dt
\end{align}
Here we are normalizing the convolution by the time width, $\tau$. Since $\sigma_{vR}\left(x\right)$ has unit area, this scaling allows $\widehat{V}^{\sigma_{vR}}$ to be interpreted as a smoothing of $V$. Application of the $L^2$ norm\footnote{We scale the integration in the $L^2$ norm by $1/{\abs{T}}$ to make the metric \emph{intensive} (see below).} to smoothed potentials gives us a generalization of van Rossum's spike train metric to membrane potential recordings and an example of a \emph{convolution metric}\footnote{Square brackets in the definition for $d_{vR}$ (Eq.~(\ref{Eq:dVanRossum})) and throughout this article indicate that the symbol being defined is a functional, having some arguments that are functions.}:
\begin{align}
\label{Eq:dVanRossum}
d_{vR}\left[V_1, V_2; \tau\right]_T & \equiv \sqrt{ \int_{-\infty}^{\infty} 
	\left( \widehat{V}^{\sigma_{vR}}_1\left(t;\tau\right)_T
	 - \widehat{V}^{\sigma_{vR}}_2\left(t;\tau\right)_T \right)^2 dt/ \! \abs{T}} \\
	& = \sqrt{ \int_{-\infty}^{\infty}
	\left( \int_T \sigma_{vR} \left(\left(t - t' \right)/ \tau\right)
	\cdot \left(V_1 \! \left(t'\right) - V_2 \! \left(t'\right)\right) dt'/\tau \right)^2 dt/ \! \abs{T}}
\end{align}

The distinction in notation is important here. $D^2_{vR}$ refers to van Rossum's original distance squared which is defined on spike trains (Eq.~(\ref{Def:D2vR})); $d_{vR}$ is defined for membrane potential recordings and is scaled differently. Like van Rossum's spike series metric, $d_{vR}^2$ compares nearby spike times in the recordings by virtue of the convolution operation's spreading of spikes in time by $\tau$. The relationship between $D^2_{vR}$ and $d_{vR}$ is:
\begin{equation}
\label{Eq:D2vRdvR}
D^2_{vR}\left(s_1,s_2\right)_{\tau} = \tau \! \abs{T} \,
	d_{vR}\left[f_{s_1} \! \left(t\right),f_{s_2} \! \left(t\right);\tau \right]_T^{\, 2}
\end{equation}

For convenience, we are treating membrane potential recordings here as continuous functions of time, neglecting the fact that actual recordings are discretely sampled. This will be the approach taken throughout the article. Toward the end, we will address approximation methods for applying the convolution metric, $d_C$, to sampled time series.

The $d_{vR}$ metric, Eq.~(\ref{Eq:dVanRossum}), has a couple of issues. First, just as van Rossum needed to take $D^2_{vR}$ in order to get a spike train distance that is first order in spike time differences, one must take $d_{vR}^2$ in order to get the same property. The first-order property provides those interested in spike timing with an advantage: the magnitude of timing differences small compared to the timing sensitivity, $\tau$, correlates well with the magnitude of the metric. For analysts interested in both spike timing and membrane potential differences, a metric that is first order in both offers a similar advantage. However, $d_{vR}^2$ is second order in $V_1\!-\!V_2$. Secondly, the infinite tail of the exponential decay in $\sigma_{vR}$ causes the hassle of needing to perform an integral, or an approximation thereof, over all recorded times prior to $t$ when evaluating the interior integral in Eq.~(\ref{Eq:dVanRossum}).

These two observations make it worthwhile that, in constructing a membrane potential metric, we attempt to replace $\sigma_{vR}$ in Eq.~(\ref{Eq:dVanRossum}) with a kernel that changes these two properties for the metric. In addition, the kernel ought to produce a favorable response for the metric to non-spiking membrane potential dynamics. In this respect, while some neglect of high-frequency components is unavoidable due to the fact that we are convolving with a continuous function, which is a smoothing operation, we want to preserve as much high-frequency information in the time course as we can. Otherwise, we want the frequency response to lack any bias with regard to specific frequency bands. Finally, we want a kernel that causes the metric to behave like a metric, viz., to give 0 if and only if the recordings it evaluates are identical. We will find a kernel (Eq.~(\ref{Def:sigC})) that addresses each of these issues, namely one that:
\begin{enumerate}
\item yields a metric with a first order response to same-time membrane potential differences and nearby spike time differences for otherwise identical recordings
\item is symmetric and non-zero only over a finite domain
\item produces a metric with a smooth low-pass frequency response that lacks any zeros, minima or oscillations which would unfairly bias or overlook certain frequencies
\item gives the metric a near-optimally fat-tailed frequency response curve.
\item causes the metric to return 0 if and only if recordings are identical
\end{enumerate}

\subsection{A generalized convolution metric}

For the purpose of analysis, we will now define a generalized convolution metric, $d_{gen}$, which accepts a smoothing kernel, $\sigma$, as one of its arguments: We will replace the van Rossum kernel, $\sigma_{vR}$, in the definition for $d_{vR}$ (Eq.~(\ref{Eq:dVanRossum})) with the kernel, $\sigma$, supplied as argument to $d_{gen}$. This construction will allow us to study the dependence of a convolution metric on its kernel function:
\begin{align}
&d_{gen}\left[\sigma\right]\left[V_1, V_2;\tau\right]_T \equiv \notag \\
&\qquad N \! \left[\sigma\right] \sqrt{\int_{-\infty}^{\infty} 
	\left(\int_T \sigma \! \left(\left(t - t' \right)/ \tau\right) 
	\cdot \left(V_1 \! \left(t'\right) - V_2 \! \left(t'\right)\right)
	 dt'/\tau\right)^2 dt/ \! \abs{T}} \label{Def:dGen} \\
&\qquad =N \! \left[\sigma\right] \sqrt{\int_{-\infty}^{\infty} 
	\left( \widehat{V}^{\sigma}_{1} \! \left( t;\tau \right)_T - 
	\widehat{V}^{\sigma}_{2} \! \left( t;\tau \right)_T \right)^2 dt/ \! \abs{T} } \label{Eq:dGenVsmooth}
\end{align}
Here, $T$ is the time domain of the membrane potential recordings; $\tau$ is the range of time over which feature (e.g., spike) timing is compared; and $\widehat{V}^{\sigma}_1\left(t;\tau\right)_T$ and $\widehat{V}^{\sigma}_2\left(t;\tau\right)_T$ are $\sigma$-smoothings (in practice, $\sigma \! \left(x\right)$ should be normalized to unit area) of $V_1$ and $V_2$: 
\begin{equation}
\widehat{V}^{\sigma}_i \! \left(t;\tau\right)_T \equiv 
	\frac1{\tau} \int_T \sigma \! \left(\left(t - t'\right)/\tau\right)V_i \! \left(t'\right) dt'
\end{equation}

We have scaled both internal and external integrals in Eq.~(\ref{Def:dGen}) by the size of the relevant time domain to produce an intensive metric. An ``intensive'' metric indicates that if two signals preserve the same pattern of difference over an extended length of time, the distance between them is the same regardless of how long the recordings are (provided $\abs{T} \gg \tau$ so that edge effects are negligible). The units of the metric are therefore Volts, not Volt-seconds. The exterior integral is scaled by $\abs{T}^{-1}$ since $\abs{T}$ is the size of the domain over which $V_1$ and $V_2$ contribute. The interior integral is scaled by $\tau^{-1}$ since the area of $\sigma \! \left(t/\tau\right)$, will be proportional to $\tau$, the smoothing time.

$N \! \left[\sigma\right]$ is a kernel-dependent normalization term. The constraint used to set $N \! \left[\sigma\right]$ will be explained in Sec.~\ref{Sec:SingleSpikes} and is a regularization of the initial rate of increase for $d_{gen}$ with spike time differences. The normalization we will arrive at is $N \! \left[\sigma\right] = 1/\sqrt{\int_{-\infty}^{\infty}\sigma' \! \left(x\right)^2dx}$, where $\sigma' \! \left(x\right)$ is the derivative of $\sigma$ with respect to its argument, assuming $\sigma' \! \left(x\right)$ is square-integrable\footnote{Note we have changed variables from $t$ to $x \equiv t/\tau$.}. For most kernels we will discuss, this latter condition holds. It does not hold for $\sigma_{vR}$; the initial rate of increase for $d_{gen}\left[\sigma_{vR}\right]$ with spike time differences diverges, and our usual constraint cannot be met. Nonetheless, we must define $N \! \left[\sigma_{vR}\right]$ in order for $d_{gen}\left[\sigma_{vR}\right]$ to be defined. In general, we use $N \! \left[\sigma\right]=1$ when $\sigma' \! \left(x\right)$ is non-square-integrable. This gives $d_{vR}=d_{gen}\left[\sigma_{vR}\right]$. Eq.~(\ref{Def:NSig}) states this definition for $N \! \left[\sigma\right]$.

\subsection{Overview}

We will begin our analysis by addressing $d_{gen}$'s response to action potentials, modeling membrane potentials as sums over Dirac delta function spikes. This will allow us to obtain the condition on $\sigma$ yielding first order dependence for $d_{gen}$ on individual spike time differences along with the normalization, $N \! \left[\sigma\right]$ (both of which were just stated). We will see that, provided the kernel meets the condition, convolution metrics respond to spike timing differences which are small compared to $\tau$ in the same way that the Euclidean metric on coordinate spaces responds to individual coordinate differences. We will discuss other aspects of a convolution metric's response to spike timing as well.

Turning our attention to $d_{gen}$'s response to non-spiking features of the membrane potential, we then benefit from some Fourier analysis. We will show that with the choice, $\sigma_X$ (Eq.~(\ref{Def:sigC})), for the kernel, we get a near-ideally-gradual low-pass frequency response that is otherwise impartial to specific frequency bands. We will continue by showing that this response implies $d_C \equiv d_{gen}\left[\sigma_X\right]$ returns zero only for identical membrane potentials. The triangle inequality for $d_C$ is shown in App.~\ref{Sec:TriangleInequality}.

Finally, we will evaluate $d_C$, $D^2_{vR}$, and $D_{V\!P}$ numerically for several types of extended neural data, comparing the different metrics' performances. The inputs to the metrics will be pairs of data, one of which is a systematically modified version of the other. The data will include randomly generated delta function Poisson spike trains and simulations of a Hodgkin-Huxley neuron under randomly generated current input. We will scale the spike train metrics, $D^2_{vR}$ and $D_{V\!P}$, in such a way that they can be plotted on the same graph with $d_C$, and the values can be directly compared with each other. In each case, we will plot $d_C$ versus a parameter that controls the timing offset for features in the recordings. We will see that the convolution metric, $d_C$, gives sensible output and verify that it provides the analyst with a considerable amount of additional information about the difference between two neural signals than what is available from either the van Rossum or Victor-Purpura spike train metrics.

\section{Results}
\label{Sec:ActionPotDeltaFn}

Let us consider the response $d_{gen} \! \left[\sigma\right]$ has to spike timing. Given two spike trains, $s_1$ and $s_2$, as defined in Eq.~(\ref{Def:SpTrain}), we will model their corresponding membrane potential time courses by treating the spikes as Dirac delta functions which capture, in idealized fashion, the large-amplitude, narrow-timescale character of the action potential:
\begin{equation}
\label{Def:V1V2Sp}
V_{s_1} \! \left(t;\alpha \right) = \alpha \! \sum_{i=1}^{n_1} \delta \! \left(t - t_{1i}\right), \quad V_{s_2} \! \left(t;\alpha\right) = \alpha \! \sum_{i=1}^{n_2} \delta \! \left(t - t_{2i}\right)
\end{equation}

The parameter, $\alpha$, sets the area under each spike; $\alpha \! \sim \! 100 \ \mu$V$\cdot \,$s is realistic. We will assume that all spikes occur within the time domain of comparison, $T$.

\subsection{Single spike recordings}
\label{Sec:SingleSpikes}

In the case of single spike time courses, $V_{s_1} \! \left(t;\alpha \right) = \alpha \ \delta \!\left(t - t_s\right), V_{s_2} \! \left(t;\alpha \right) = \alpha \ \delta \! \left(t - \left(t_s + t_d\right)\right)$, with spikes occurring $t_d$ apart, $d_{gen}$ evaluates as:
\begin{align}
& d_{gen} \left[\sigma\right] \left[V_{s_1} \! \left(t;\alpha \right),
	V_{s_2} \! \left(t;\alpha \right);\tau\right]_T = \notag \\
& \qquad \qquad \phantom{= .} N \! \left[\sigma\right] \sqrt{\int_{-\infty}^{\infty} \frac{\alpha^2}{\tau^2} 
	\left(\sigma \! \left(\left(t - t_s\right)/\tau\right) 
	- \sigma \! \left(\left(t - t_s - t_d\right)/ \tau\right)\right)^2 dt / \! \abs{T}} \\
& \qquad \qquad = N \! \left[\sigma\right] \frac{\alpha \sqrt2}{\sqrt{ \tau \abs{T}}}
	 \ \sqrt{\left( \int_{-\infty}^{\infty} \sigma \! \left(x\right)^2 dx - 
	\int_{-\infty}^{\infty} \sigma \! \left(x\right)\sigma \! \left(x-t_d/\tau\right)dx\right)} 
\label{Eq:dGenSingSpikes}
\end{align}

Here we have changed variables from ``real time,'' $t$, measured in seconds, to dimensionless ``kernel time,'' $x \equiv t/\tau$. From here on, we will be frequently switching back and forth between these as is convenient. We see that the $d_{gen}$ metric is proportional the square root of $\sigma$'s autocorrelation at zero lag minus its autocorrelation at lag $t_d/\tau$.

Adopting the notation $R \! \left[\sigma,\sigma\right]\left(x\right)$ for the autocorrelation, we may write:
\begin{align}
& d_{gen}\left[\sigma\right] \left[V_{s_1} \! \left(t;\alpha \right),
	V_{s_2} \! \left(t;\alpha \right);\tau\right]_T 
	 =\tfrac{\alpha}{\sqrt{\tau \abs{T}}} \ \gamma_{gen} \left[\sigma\right] \left(t_d/\tau \right) \\
\label{Def:GammaGen}
& \mbox{where } \gamma_{gen}\left[\sigma\right] \left(x\right) \equiv 
	N \! \left[\sigma\right] \sqrt{ 2\left( R \! \left[\sigma,\sigma\right]\left(0\right) 
	- R \! \left[\sigma,\sigma\right]\left(x\right) \right) }
\end{align}
for single spike time recordings.

Noting that the autocorrelation for $\sigma_{vR}\left(x\right)$ is:
\begin{align}
R\left[\sigma_{vR},\sigma_{vR}\right]\left(x\right) & \equiv
	\int_{-\infty}^{\infty}  \sigma_{vR}\left(x'\right)\sigma_{vR}\left(x' - x\right)dx' 
	= \tfrac12 e^{-\abs{x}} \, ,
\end{align}
and recalling $N\! \left[\sigma_{vR}\right] \equiv 1$, we have for single spikes:
\begin{align}
& d_{vR}\left[V_{s_1} \! \left(t;\alpha \right),
	V_{s_2} \! \left(t;\alpha \right);\tau\right]_T  = 
	d_{gen}\left[\sigma_{vR}\right]
	\left[V_{s_1} \! \left(t;\alpha \right),V_{s_2} \! \left(t;\alpha \right);\tau\right]_T =
	\tfrac{\alpha}{\sqrt{\tau \abs{T}}} 
	\, \sqrt{1 - e^{-\abs{t_d}/\tau}} 	\label{Eq:dvRss} \\	
& \Rightarrow d_{vR}\left[V_{s_1} \! \left(t;\alpha \right),
	V_{s_2} \! \left(t;\alpha \right);\tau\right]_T^{\, 2}  = 
	\tfrac{\alpha^2}{\tau \abs{T}}  \left(1 - e^{-\abs{t_d}/\tau} \right)
	\label{Eq:dvR2ss}
\end{align}
Since $f_{s_i} \! \left(t\right) \! = \! V_{s_i} \! \left(t;1.0\right)$ (Eqs.~\ref{Def:fs}, \ref{Def:V1V2Sp}) and $D^2_{vR}\left(s_1,s_2\right)_\tau = \tau \! \abs{T} \, d_{vR}\left[f_{s_1},f_{s_2};\tau\right]_T^{\, 2}$ (Eq.~(\ref{Eq:D2vRdvR})), this is consistent with van Rossum's result (Eq.~(\ref{d2vR_ss})).

With the formula, Eq.~(\ref{Eq:dvRss}), we see the problem with $d_{vR}$ and $D_{vR}$ we previously discussed in Secs.~\ref{Sec:Bg} and \ref{Sec:Methods}: the initial rise is in proportion to $\sqrt{\abs{t_d}}$. It is therefore necessary to square, as van Rossum and others have, to get a quantity that initially rises as $\abs{t_d}$. But in our case this would produce a metric that is second order in $V_{s_1}-V_{s_2}$, with dimensions of membrane potential squared.

We can avoid having to do this by placing a proper constraint on the kernel, $\sigma$, namely that $\int_{-\infty}^{\infty} \sigma' \! \left(x\right)^2 dx$ converges to a non-zero finite value. To see that this suffices, we first express $\gamma_{gen}$ as follows (letting $x_d \equiv t_d/\tau$):
\begin{align}
	\gamma_{gen}\left[\sigma\right]\left( x_d \right) & = N \! \left[\sigma\right] 
	\sqrt{\int_{-\infty}^{\infty} \left(\sigma \! \left(x\right) - \sigma \! \left(x + x_d\right)\right)^2 dx} 
	\label{Eq:GammaGenSS_sig} 
\end{align}

We may examine the behavior of $\gamma_{gen}$ for $x_d$ close to zero by Taylor expanding $\sigma \! \left(x + x_d\right)$ in the integrand in powers of $x_d$ about $x$, which gives:
\begin{align}
&\gamma_{gen}\left[\sigma\right]\left(x_d\right) = N \! \left[\sigma\right]
	\sqrt{\int_{-\infty}^{\infty} x_d^2 \cdot \left( 
	\begin{aligned}
	& \sigma' \! \left(x\right) + \tfrac12  \sigma'' \! \left(x\right) \cdot x_d + \ldots
	\end{aligned}  \right)^2 dx} \label{Eq:dgen_ss-Texp}
\end{align}

Since we are taking the limit $x_d \to 0$, we will dismiss all but the lowest order term in $x_d$ in the integrand to give (recalling $x_d = t_d/\tau$):
\begin{align}
&\lim_{t_d \to 0} \gamma_{gen}\left[\sigma\right]\left(t_d/\tau\right)
	= \frac{\abs{t_d}}{\tau} N \! \left[\sigma\right] \sqrt{\int_{-\infty}^{\infty} \sigma' \! \left(x\right)^2 dx} 
	\label{Eq:InitRiseGGl}
\end{align}
This dismissal may raise some doubt since $\sigma''$ can involve divergences, as it will in case of $\sigma_X$. It may therefore seem questionable whether terms involving $\sigma''$ and higher derivatives can be reliably neglected as $x_d \to 0$ even though they involve higher powers of $x_d$. As it turns out, so long as $\int_{-\infty}^{\infty} \sigma' \! \left(x\right)^2 dx$ exists, Eq.~(\ref{Eq:InitRiseGGl}) is valid. This is rigorously established in Appendix~\ref{Appdx:dglRise}.

Our sufficient condition, then, for a proportional rise of $d_{gen}$ with the spike time difference between single delta-function-spike recordings is that:
\begin{equation}
\label{Cond:SigOK}
0 < \int_{-\infty}^{\infty} \sigma' \! \left(x\right)^2 dx < \infty
\end{equation}
Eq.~(\ref{Eq:InitRiseGGl}) is also our basis for setting $N \! \left[\sigma\right]$. For any kernel satisfying Eq.~(\ref{Cond:SigOK}), we ensure that $\gamma_{gen}$ initially rises as $\slfrac{\abs{t_d}}{\tau}$ (and that $d_{gen}$ therefore rises as $\frac{\alpha}{\tau^{\nicefrac32} \sqrt{\abs{T}}} \abs{t_d}$) if we set $N \! \left[\sigma\right] = \slfrac1{\sqrt{\int_{-\infty}^{\infty} \sigma' \! \left(x\right)^2 dx }}$. This is a useful way to regularize the behavior of $d_{gen}$, and we use it. For any $\sigma$ that does not satisfy Eq.~\ref{Cond:SigOK} (in particular, $\sigma_{vR}$), we do not attempt any regularization and set $N \! \left[\sigma\right]$ to unity:
\begin{align}
N \! \left[\sigma\right] = 
& \begin{cases}
\slfrac{1}{\sqrt{\int_{-\infty}^{\infty} \sigma' \! \left(x\right)^2 dx}}
	 \ & : \ 0 < \int_{-\infty}^{\infty} \sigma' \! \left(x\right)^2 dx < \infty \\
1 & : \ \mbox{otherwise}
\end{cases} 
\label{Def:NSig}
\end{align}

\subsection{Many-spike recordings}
\label{Sec:MnySpkRec}

Let us now move on to uncover how $d_{gen}$ performs when many spikes are involved:
\begin{align}
V_{s_1} \! \left(t;\alpha \right) =
	\alpha \! \sum_{i=1}^{n_1} \delta \! \left(t - t_{1i}\right) \, , \ 
	V_{s_2} \! \left(t;\alpha \right) = 
	\alpha \! \sum_{i=1}^{n_2} \delta \! \left(t - t_{2i}\right)
\end{align}

For such time courses, we have:
\begin{align}	
&d_{gen}\left[\sigma\right] \left[V_{s_1} \! \left(t;\alpha \right), 
	V_{s_2} \! \left(t;\alpha \right);\tau\right]_T = \notag \\
& \quad \phantom{ 1 = } \frac{N \! \left[\sigma\right] \alpha}{\tau} 
	\sqrt{\int_{-\infty}^{\infty} \left(\sum_{i=1}^{n_1} 
	\sigma \! \left(\left(t - t_{1i}\right)/\tau\right) - \sum_{i=1}^{n_2}  \sigma \! \left(\left(t - t_{2i}\right)/ \tau\right)\right)^2 dt / \! \abs{T}} \label{Eq:dGenMany1} \\ \notag \\
&\quad = \frac{N \! \left[\sigma\right] \alpha}{\sqrt{\tau \abs{T}}} \sqrt{
	 \begin{aligned} 
	 & \left(n_1 + n_2\right)R \! \left[\sigma,\sigma\right]\left(0\right) + 2\sum_{i=1}^{n_1} 
	 \sum_{j=1}^{i-1}R \! \left[\sigma,\sigma\right]\left(\frac{t_{1i} - t_{1j}}{\tau}\right) \\
	&+ 2\sum_{i=1}^{n_2} \sum_{j=1}^{i-1}R \! \left[\sigma,\sigma\right]
	\left(\frac{t_{2i} - t_{2j}}{\tau}\right) 
	-2\sum_{i=1}^{n_1} \sum_{j=1}^{n_2} R \! \left[\sigma,\sigma\right]
	\left(\frac{t_{1i} - t_{2j}}{\tau}\right)
	\end{aligned}} \label{Eq:dGenManySpikes}
\end{align}

Note that for compactly supported $\sigma$, the sums in Eq.~(\ref{Eq:dGenManySpikes}) only need to be explicitly evaluated over terms for which the scaled time difference, $(t_a - t_b)/\tau$, is less than some critical value---beyond which $R \! \left[\sigma,\sigma\right]$ will be identically zero. Assuming a density of spikes that is, on average, constant in time, this leads to computational complexity for the exact calculation that grows as $n_i + n_j$.  Conversely, for non-compact $\sigma$ such as $\sigma_{vR}$, this truncation is not possible, leading to $O \! \left( n_i \! \cdot \! n_j \right)$ complexity as one has for $D^2_{vR}$.

Because we have set $N \! \left[\sigma_{vR}\right]=1$,
\begin{equation} d_{vR}\left[V_{s_1} \! \left(t;\alpha \right),V_{s_2} \! \left(t;\alpha \right);\tau\right] 
	= d_{gen}\left[\sigma_{vR}\right] \left[V_{s_1} \! \left(t;\alpha \right),
		V_{s_2} \! \left(t;\alpha \right);\tau\right]
\end{equation}

Recall that
\begin{equation}
R\left[\sigma_{vR},\sigma_{vR}\right]\left(t/\tau\right) =
	\tfrac12 e^{-\abs{t}/\tau}
\end{equation}
Since $D^2_{vR}\left(s_1,s_2\right)_\tau = \tau \! \abs{T} \, d_{vR}\left[V_{s_1},V_{s_2};\tau\right]^2$, Eq.~(\ref{Eq:dGenManySpikes}) is consistent with Eq.~(\ref{Eq:D2vRFar}), and both are consistent with an increasing $d_{vR}$ as spikes in opposite trains separate.

In particular, for single spike trains, both expressions recover Eq.~(\ref{d2vR_ss}):
\begin{equation}
D^2_{vR}\left(s_1,s_2\right)_\tau = \tau \! \abs{T} \, 
	d_{vR}\left[V_{s_1} \! \left(t;\alpha \right),
	V_{s_2} \! \left(t;\alpha \right);\tau\right]^2 = 1 - e^{-\abs{t_d}/\tau} \, ,
\end{equation} where $t_d$ is the spike time difference. This function is inversely peaked at $t_d=0$ and is identical to \citeauthor{vRossum}'s \citeyearpar{vRossum} expression.

We can clarify Eq.~(\ref{Eq:dGenManySpikes}) by making use of $\gamma_{gen}\left[\sigma\right]\left(x\right)$ (Eq.~(\ref{Def:GammaGen})):
\begin{align}
&d_{gen}\left[\sigma\right] \left[V_{s_1} \! \left(t;\alpha \right),
	V_{s_2} \! \left(t;\alpha \right);\tau\right]_T =  \notag \\
&\qquad \frac{\alpha}{\sqrt{\tau \abs{T}}} \sqrt{
	 \begin{aligned} 
	 & \left(n_1 - n_2\right)^2 N \! \left[\sigma\right]^2 R \! \left[\sigma,\sigma\right]\left(0\right)
	- \sum_{i=1}^{n_1} \sum_{j=1}^{i-1} \gamma_{gen}\left[\sigma\right]\left(\left(t_{1i} - t_{1j}\right)/\tau\right)^2\\
	&- \sum_{i=1}^{n_2} \sum_{j=1}^{i-1} \gamma_{gen}\left[\sigma\right]\left(\left(t_{2i} - t_{2j}\right)/\tau\right)^2
	+ \sum_{i=1}^{n_1} \sum_{j=1}^{n_2} \gamma_{gen}\left[\sigma\right]\left(\left(t_{1i} - t_{2j}\right)/\tau\right)^2
	\label{Eq:dStri_dss}
	\end{aligned}}
\end{align}

Assuming the autocorrelation, $R \! \left[\sigma,\sigma\right]\left(x\right)$ vanishes as $x$ goes to $\infty$, which it does for $\sigma_{vR}$, along with any $\sigma$ with compact support, we have:
\begin{align}
&d_{gen}\left[\sigma\right] \left[V_{s_1} \! \left(t;\alpha \right),
	V_{s_2} \! \left(t;\alpha \right);\tau\right]_T =  \notag \\
&\qquad \frac{\alpha}{\sqrt{\tau \abs{T}}} \sqrt{
	 \begin{aligned} 
	 & \nicefrac12 \left(n_1 - n_2\right)^2 \gamma_{gen}\left[\sigma\right]\left(\infty\right)^2
	- \sum_{i=1}^{n_1} \sum_{j=1}^{i-1} \gamma_{gen}\left[\sigma\right]\left(\left(t_{1i} - t_{1j}\right)/\tau\right)^2\\
	&- \sum_{i=1}^{n_2} \sum_{j=1}^{i-1} \gamma_{gen}\left[\sigma\right]\left(\left(t_{2i} - t_{2j}\right)/\tau\right)^2
	+ \sum_{i=1}^{n_1} \sum_{j=1}^{n_2} \gamma_{gen}\left[\sigma\right]\left(\left(t_{1i} - t_{2j}\right)/\tau\right)^2
	\label{Eq:dGenMultSpk}
	\end{aligned}}
\end{align}
where $\gamma_{gen}\left[\sigma\right]\left(\infty\right)$ is shorthand for $\lim_{x \to \infty} \gamma_{gen}\left[\sigma\right]\left(x\right)$. Finitude of $\gamma_{gen}\left[\sigma\right]\left(\infty\right)$ is guaranteed if we further assume square-integrability for $\sigma$ since if $R \! \left[\sigma,\sigma\right]\left(x\right)$ vanishes with large $x$, $\gamma_{gen}\left[\sigma\right]\left(\infty\right)=\sqrt{2 \, R \! \left[\sigma,\sigma\right]\left(0\right)}=\sqrt{2 \int_{-\infty}^{\infty} \sigma \! \left(x\right)^2 dx}$. Again, terms only need to be explicitly summed over if $\nicefrac{t_a - t_b}{\tau}$ is not so large that $\gamma_{gen} \left[\sigma\right] \left( \nicefrac{t_a - t_b}{\tau}\right) = \gamma_{gen}\left[\sigma\right]\left(\infty\right)$. For compactly supported $\sigma$, this often allows an $O \! \left(n_i + n_j\right)$ complexity calculation.

Eq.~(\ref{Eq:dGenMultSpk}) allows us to better see the inner workings of the metric: $d_{gen}^2$ has a positive term proportional to the squared difference in the number of spikes for the two recordings, and it also has a positive term which increases with the time difference between pairs of spikes, one from each recording, in precisely the same functional manner as the metric does for a single pair of spikes. In particular, the term will increase in the time difference, initially, as $\left(\left(t_{1i}-t_{2j}\right)/\tau\right)^2$, eventually approaching an upper bound as the difference sufficiently exceeds $\tau$. This leads to an initially hyperbolic rise for $d_{gen}$ with individual time differences (linear if the membrane potentials are otherwise equal). The parameter, $\tau$, sets the scale for the distance's (bounded) dependence on spike time differences. The hyperbolic behavior is favorable because it is the same functional response that the Euclidean metric on coordinate spaces has with individual coordinate differences.

The two positive terms make a great deal of sense. Pairs of spikes between recordings are compared in a sensible way, and the presence of unpaired spikes raises the distance squared as the square of the number of such spikes. Interestingly two other terms appear in the distance squared, both of which \emph{decrease} with the time difference between pairs of spikes in the \emph{same} recording, according to the exact same function as does the distance squared \emph{increase} in the time difference between pairs in opposite recordings. These terms may be compared to the time-dependence on pairs of same-train spikes in the first two double sums in Eq.~(\ref{dvR2_PaivaEtAl}). We know, from $d_{gen}$'s definition (Eq.~(\ref{Def:dGen})), that it is strictly positive and that these negative terms cannot overcome the positive ones. Furthermore, their presence can be anticipated from the form of Eq.~(\ref{Eq:dGenMany1}).

With these recognitions quieting any doubts, it is useful to remark on the utility of the negative terms, which may be seen as handling the problem of spike pairing: Given two spike trains, one of which differs from the other only by small timing shifts in the same spikes, it is natural to think that only the timing differences between the spikes that ``go together'' is important to the distance between the trains. The relative timing of unrelated spikes is not so important. The negative terms address this issue by causing convolution metrics to neglect differences in less related spike pairs and to focus on differences in more related pairs---in the following way: Suppose two spikes, $t_{1i}$ and $t_{2j}$, in opposite recordings, are nearby to one another, and that a third spike, say $t_{1k}$, is far from both. Our $d_{gen}$ prioritizes the timing difference between $t_{1i}$ and $t_{2j}$ by ``shielding'' the effect of the positive inter-recording term, $\gamma_{gen}\left[\sigma\right]\left(\left(t_{1k}-t_{2j}\right)/\tau\right)^2$, with the negative intra-recording term, $-\gamma_{gen}\left[\sigma\right]\left(\left(t_{1k}-t_{1j}\right)/\tau\right)^2$. Due to the asymptotic behavior of $\gamma_{gen}$ (for well-behaved $\sigma$), these terms will be similar in magnitude.

\subsection{Fourier analysis}
\label{Sec:dGlFSpace}

We now need to address how our metric handles oscillatory components of various frequencies in the input time courses. Since it looks at smoothed membrane potentials, $d_{gen}$ will, by necessity, have an attenuated response to high frequencies. This is by design; we want the metric to deemphasize the highest frequency details in the recordings, viz., the non-overlap of action potentials that occur reasonably near one another in time. For sub-spike membrane potential fluctuations, a low-pass response is okay, though we would usually like to limit this characteristic as much as possible\footnote{In some cases, such as in the presence of abundant high-frequency noise, this may not apply. This circumstance is addressed in App.~\ref{App:sigNu}}. There should be no frequencies, however, which are \emph{entirely} overlooked. Furthermore, we don't want any local minima or maxima in the frequency response other than the central peak\footnote{The tip of this peak will usually not contribute substantially since we are taking the \emph{difference} between membrane potential recordings, which tends toward zero mean for sufficiently lengthy signals; the primary contribution to the metric will come from the slopes of the central response peak.} at $\omega=0$, since this would be unjustifiably partial toward or against such frequencies.

Recall our definition of $d_{gen}$:
\begin{align}
&d_{gen}\left[\sigma\right]\left[V_1, V_2;\tau\right]_T = \notag \\
&\quad N \! \left[\sigma\right] \sqrt{\int_{-\infty}^{\infty} \left(\int_T \sigma \! \left(\left(t - t' \right)/ \tau\right) \cdot \left(V_1 \! \left(t'\right) - V_2 \! \left(t'\right)\right) dt'/\tau \right)^2 dt/ \! \abs{T}} \end{align}

We may alternatively express this in the following way:
\begin{align}
& d_{gen}\left[\sigma\right]\left[V_1, V_2; \tau\right]_T = N \! \left[\sigma\right] \sqrt{\int_{-\infty}^{\infty} s \! \left[\sigma;\tau\right]\left(t\right)_T^{\, 2} \, dt/ \! \abs{T} } \label{Eq:dGens} \\
& \mbox{where } I \! \left(t\right)_T \equiv 
	\begin{cases}
		1 &: t \in T  \\
		0 &: \mbox{ otherwise}
	\end{cases} ; \\
& \phantom{\mbox{where }} h \! \left(t\right)_T \equiv I \! \left(t\right)_T \! \cdot \left(V_1 \! \left(t\right) - V_2 \! \left(t\right)\right) ;
	\label{Def:hT} \\
& \phantom{\mbox{where }} \mbox{and } s \! \left[\sigma;\tau\right]\left(t\right)_T \equiv \int_{-\infty}^{\infty} \sigma \! \left(\left(t-t'\right)/\tau\right) h \! \left(t'\right)_T \, dt'/\tau \label{Def:s}
\end{align}

Parseval's theorem informs us that Eq.~(\ref{Eq:dGens}) is equivalent to an integral over the square modulus of the Fourier transform, $\tilde{s} \! \left(\omega\right)$, of $s \! \left(t\right)$:
\begin{align}
d_{gen}\left[\sigma;\tau\right]\left[V_1, V_2\right]_T &= \frac{N \! \left[\sigma\right]}{\sqrt{\abs{T}}}  \sqrt{\int_{-\infty}^{\infty} \abs{\tilde{s}\left[\sigma;\tau\right]\left(\omega\right)_T}^2  d \omega }
\end{align}

From the convolution theorem for Fourier transforms, we have that:
\begin{align}
\tilde{s} \left[\sigma;\tau\right] \left( \omega \right)_T &= \tilde{\sigma} \! \left(\tau \omega\right) \tilde{h} \! \left( \omega \right)_T
\end{align} 

Therefore,
\begin{equation}
d_{gen}\left[\sigma;\tau\right]\left[V_1, V_2\right]_T = 
	\frac{N \! \left[\sigma\right]}{\sqrt{\abs{T}}}  
	\sqrt{\int_{-\infty}^{\infty} \abs{\tilde{\sigma} \! \left(\tau \omega\right)}^2 
	\cdot  \abs{\tilde{h} \! \left(\omega\right)_T}^2 d \omega } \label{Eq:DGenFspace}
\end{equation}

We see that $d_{gen}$ responds to frequencies, $\omega$, in $h$ (which will be peaked around those in $V_1 \! - \! V_2$) according to the presence of the $\tau$-scaled frequency, $\tau \omega$, in $\sigma$. Said another way, $\sigma$'s power spectrum, $\abs{\tilde{\sigma} \! \left(\tau \omega\right)}^2$, \emph{defines} the frequency response for $d_{gen}$. We can therefore ensure that the frequency response does not unfairly bias or overlook specific frequencies or frequency bands by stipulating that our kernel function's Fourier transform is a smooth, oscillation-free function that has no zeros. As we will see in Sec.~\ref{Sec:IdInd}, the no-zeros property also guarantees a metric that returns 0 only for identical recordings.

\citet{SchrauwenCampenhout2007} offer three alternative kernels for spike train metrics. Two of these, the Gaussian and Laplacian kernels, are non-compact, ruling them out for us. The third, a triangularly shaped kernel, has spectral oscillations and zeros.

In the context of spectral analysis, compactly supported multiplicative window functions are sought out which minimize spectral leakage across bins when the finite Fourier transform is taken. The usual emphasis in designing these is a narrow main lobe for the window's power spectrum and not on the elimination of zeros or oscillations occurring outside the main lobe, which are quite common and present for the reputed zeroth order prolate spheroidal wavefunction, or Slepian window.

Compactly supported windows with the no-spectral-oscillations property have been studied in the context of convergent spectral parameter estimation by \citet{DepalleHelie1997}. They considered the forms:
\begin{align}
& &&\sigma_{H \! P} \! \left(x;\alpha\right) = \tfrac12 \left(1 + \cos\left(2 \pi x\right)\right) \mathrm{e}^{-2 \, \alpha \abs{x}} &&: \abs{x} \le \nicefrac12 \ ; \quad \left(\alpha \ge 2\right) \\
& \quad \mbox{and } && && \notag \\
& && \sigma_{D \! H} \! \left(x;a,b\right) = \left( 1 - 2\abs{x}\right)^a \mathrm{e}^{-4b \, x^2} &&: \abs{x} \le \nicefrac12 \ ; \quad \mbox{(specific pairs $a$, $b$)}
\end{align}
with both functions set to zero for $\abs{x} > \nicefrac12$.

The first of these is called the Hanning-Poisson window \citep[see][]{Harris1978}. The second is due to \citeauthor{DepalleHelie1997} themselves. I advance a different kernel function in this article, $\sigma_X$. The reason is that whereas for spectral analysis, one is interested in window functions that have as much low frequency power as possible \citep{Harris1978}, for a convolution metric, the more \emph{high} frequency power, the better: We want our metric to pay as much attention to high frequency information as possible\footnote{As noted above, the inverse may apply to noisy recordings: we may want to disregard high frequency components of the signal. App.~\ref{App:sigNu} provides a kernel addressing this situation.} while still retaining its spike time comparison properties---which require $0 < \int_{-\infty}^{\infty} \sigma' \! \left(x\right)^2 dx < \infty$.

\begin{figure}
\centering
\includegraphics[scale=0.8]{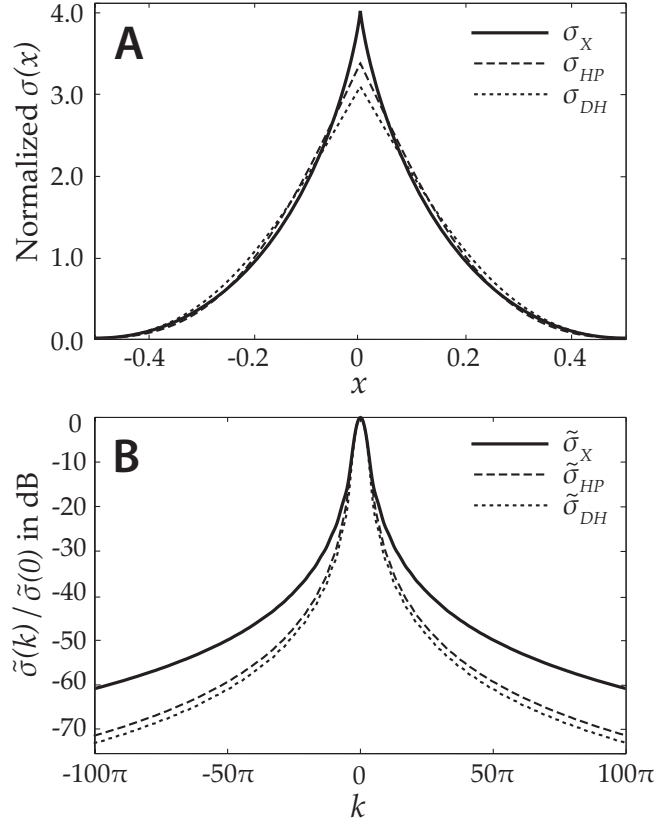}
\caption{Plots of the three kernel functions mentioned in the text that satisfy the constraints of compact support and positive, local-minimum-free frequency spectrum. Here, we are using $\alpha=2$ for $\sigma_{H \! P}$ and $a=1.8$, $b=0.92$ for $\sigma_{D \! H}$. These are the parameters for $\sigma_{H \! P}$ and $\sigma_{D \! H}$ which, of those discussed in the literature, have the greatest high-frequency response while also giving a minimum-free spectrum. In \emph{A}, the functions are plotted. For the sake of comparison, all have been normalized to unit area. In \emph{B}, we have the Fourier transforms in decibels relative to the zero-frequency amplitude.}
\label{Fig:KernelFigure}
\end{figure}

The kernel we use, $\sigma_X$, may be obtained by integrating over triangular kernels:
\begin{align}
\sigma_X \! \left(x\right)
& \equiv \begin{cases}
		2 \int_{2 \abs{x}}^1 2 \left(1 - 2\abs{x}/x_0\right)dx_0 
			= 4\left(1 + 2\abs{x}\left(\log \abs{2x} - 1\right)\right)\ &: \ 0 < \abs{x} < \nicefrac12 \\ 
		4 & : x=0 \\
		0 & : \mbox{ otherwise}
	\end{cases}
\label{Def:sigC}
\end{align}
This function is monotonically decreasing in $\abs{x}$ and is continuous and bounded at the origin. Its derivative, on the other hand, diverges at the origin, producing an infinitely sharp cusp (see Fig.~\ref{Fig:KernelFigure}). This critical feature is what causes convolution by $\sigma_X$ to preserve a near-optimal amount of the high frequency information in the signals. Nonetheless, $\int_{-\infty}^{\infty} \sigma_X'\left(x\right)^2 dx =128$ converges as required.

The Fourier transform of $\sigma_X$ is (see Fig.~\ref{Fig:KernelFigure} for plot):
\begin{equation}
\tilde{\sigma}_X \! \left(k\right) = 2 \sqrt{\tfrac2{\pi}} \left(\tfrac2{k}\right)^2 \Cin\left(k/2\right)
\end{equation}
where $\Cin\left(x\right) \equiv \int_0^{x} t^{-1}\left(1 - \cos (t)\right) dt$ ---a type of cosine integral function. The tails of $\tilde{\sigma}_X$ fall off roughly as $\log\abs{k}/k^2$. While discontinuous finite functions, such as $\sigma_{vR}$, preserve more of the high-frequency spectrum, rolling off as $\slfrac1{\abs{k}}$, they also cause $\int_{-\infty}^{\infty} \sigma' \! \left(x\right)^2 dx$ to diverge. Functions with discontinuous but finite derivatives, meanwhile, such as $\sigma_{HP}$ and $\sigma_{DH}$, roll off as $1/k^2$ \cite[see, e.g.,][pg.~59]{Harris1978}. $\sigma_X$ falls off slower without causing a divergent square integral for the derivative: it's just right for our application.

Furthermore:
\begin{align}
\tilde{\sigma}'_X \! \left(k\right) = 2 \sqrt{\tfrac2{\pi}} \left(\tfrac2{k}\right)^3
	\left( -\Cin\left(k/2\right) + 1 - \cos\left(k/2\right) \right)
\end{align}
is strictly negative for $k > 0$ and antisymmetric on $k$. This means $\tilde{\sigma}_X \! \left(k\right)$ is strictly decreasing with $\abs{k}$, as we require. 

Noting $\int_{-\infty}^{\infty} \sigma_X' \! \left(x\right)^2 dx = 128$, we will then define our preferred metric, $d_C \equiv d_{gen}\left[\sigma_X\right]$:
\begin{equation}
\label{Def:dC}
d_C\left[V_1,V_2;\tau\right]_T \equiv \frac{1}{8 \sqrt2} \sqrt{ \int_{-\infty}^{\infty} \left( \int_T \sigma_X\! \left(\left(t' - t\right)/\tau\right) \left(V_1 \! \left(t\right) - V_2 \! \left(t\right)\right) dt'/\tau \right)^2 dt/ \! \abs{T} }
\end{equation}

\begin{figure}
\centering
\includegraphics[scale=1.0]{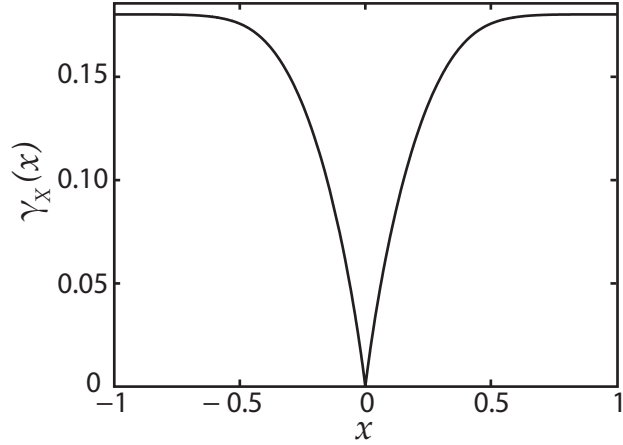}
\caption{Gamma function, $\gamma_X \! \left(x\right)$, describing the behavior of $d_C$ as spike pairs separate.}
\label{Fig:gchi}
\end{figure}

The gamma function, $\gamma_X \! \left(x\right) = \gamma_{gen}\left[\sigma_X\right]\left(x\right)$ is plotted in Fig.~\ref{Fig:gchi}.

\subsection{Identity of indiscernibles for $d_C$}
\label{Sec:IdInd}

We are now ready to show that the kernel function, $\sigma_X$ produces a metric, $d_C$, that returns 0 if and only if the two membrane potentials being compared are equal over the time domain of interest. This is evident when we consider the frequency-space expression of $d_C$ derived in the previous section (Eq. \ref{Eq:DGenFspace}):

\begin{equation}
d_C\left[V_1, V_2; \tau\right]_T = \frac1{8 \sqrt{2 \abs{T}}}  \sqrt{\int_{-\infty}^{\infty} \abs{\tilde{\sigma}_X \! \left(\tau \omega\right)}^2 \cdot  \abs{\tilde{h} \! \left(\omega\right)_T}^2 d \omega }
\end{equation}

Given that $\sigma_X$'s Fourier transform, 
\begin{equation}
\tilde{\sigma}_X \! \left(k\right) = 2\sqrt{\tfrac2{\pi}} \left(\tfrac2{k}\right)^2 \Cin\left(k/2\right), 
\end{equation}
has no zeros, $d_C$ will be zero strictly for $\tilde{h} \! \left(\omega\right)_T=0$, which will be the case if and only if $h \! \left(t\right)_T = 0$, and this is equivalent to $V_1 \! \left(t\right) = V_2 \! \left(t\right)$ over all of $T$ (see Eq. \ref{Def:hT}).

\subsection{Demonstrations}
\label{Sec:Demonstrations}

\begin{figure}
\centering
\includegraphics[scale=0.78]{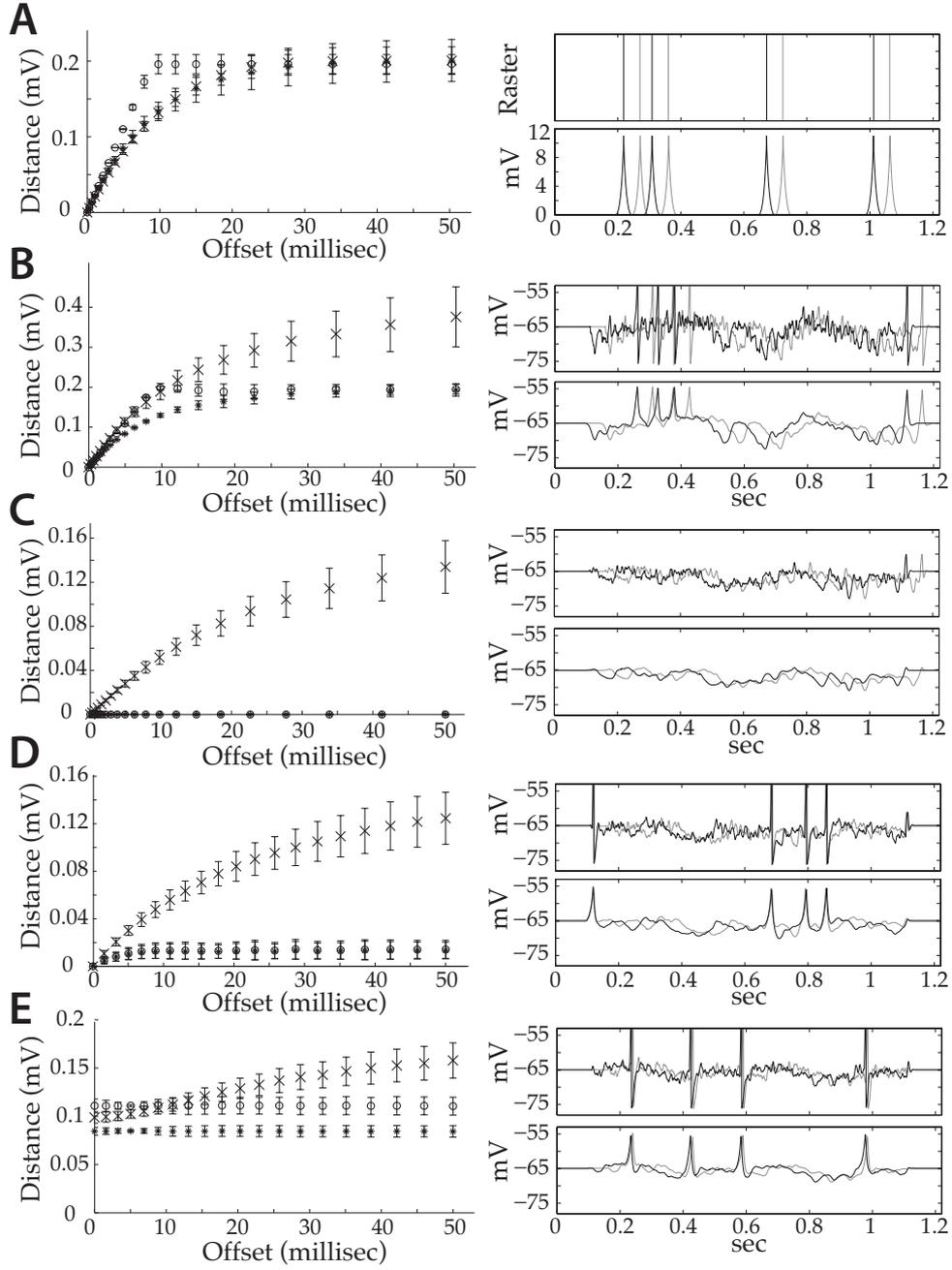}
\caption{Left column: Convolution ($\times$), scaled van Rossum ($*$), and scaled Victor-Purpura ($\circ$) metrics evaluated on several different types (\emph{A}--\emph{E}) of computer-generated neural data versus the offset of features of the data. The two arguments for the metrics are a randomly generated signal and a modified version of the same signal. See text for details. Right column: Sample raw and $\sigma_X \,$-convolved data of each type: (\emph{A}) Delta function Poisson spikes at 4 Hz. (\emph{B}) Hodgkin-Huxley neuron under inject current noise sufficient to cause 4 Hz spiking. (\emph{C}) Hodgkin-Huxley neuron under inject current noise insufficient to cause spiking. (\emph{D}) Hodgkin-Huxley neuron under input similar to \emph{C} plus stationary 4 Hz Poisson pulse input. (\emph{E}) Hodgkin-Huxley neuron under input similar to \emph{C} plus fixed (5 ms) offset 4 Hz Poisson pulse input.}
\label{Fig:metricDemos}
\end{figure}

We are now ready to compare the performance of our convolution metric, $d_C$, to that of the van Rossum metric, $D^2_{vR}$, and the Victor-Purpura metric, $D_{V \! P}$. We will do this by evaluating all three metrics with, as arguments, computer-generated neural data of several different kinds. One of the data arguments will be a systematically altered version of the other. We will plot each metric versus an ``offset'' parameter that controls the amount of alteration that occurs. The data include randomly-generated Poisson spike trains and simulated Hodgkin-Huxley neuron data (standard model parameters used---see, e.g., \citet{DayanAbbott2001}). The relevant plots are found in Fig.~\ref{Fig:metricDemos}.

In order to compare the metrics on the same plot, we choose their parameters and scale them such that, for delta function spike input, they all initially increase at the same rate with spike separation and reach the same maximum values as spikes get very far apart. Choosing a smoothing time of $\tau_X \! = \! 50$ ms for $d_C$, this requires a time constant $\tau_{vR}=\tau_X \, \gamma_X \!\left(1\right) \approx 9$ ms for $D^2_{vR}$ and a shift cost $q_{V \! P}= 2 \, c_{in}/\left(\tau_X \, \gamma_X \!\left(1\right)\right) \approx 0.22$ ms$^{-1}$ for $D_{V \! P}$, using a spike insertion/replacement cost, $c_{in}$, of 1. Leaving $d_C$ unscaled, we must scale $D^2_{vR}$ by $\alpha \, \gamma_X \! \left(1\right)/ \! \sqrt{n\abs{T} \tau_X}$, where $T$ is our time domain, $n$ is the number of spikes that occur, and $\alpha$ is the area of a spike. $D_{V \! P}$ must scale by $\alpha \, \gamma_X\! \left(1\right)/ \! \left(2 \, c_{in}\sqrt{n \abs{T }\tau_X}\right)$.

Performing these scalings requires an estimate of the spike area, $\alpha$, for simulated neurons. In App.~\ref{App:alphaEstimate} the procedure for producing this estimate is provided.

Fig.~\ref{Fig:metricDemos}(A) shows the metrics evaluated for 1-second long 4 Hz Poisson spike trains (generated by randomly inserting four spikes into a 1-second block of time). The delta function spike model is used as membrane potential input to $d_C$ with $\alpha$ set to\footnote{This value was chosen because it is consistent with the effective spike area found for the Hodgkin-Huxley neuron simulated in making the remaining four plots.} 138 $\mu$V$\cdot \,$s. Each point in the graph on the left represents a metric distance between two versions of the same randomly generated Poisson spike train: the original and a copy offset by the amount of time indicated on the $x$-axis. To keep the endpoints of the recordings fixed, 0.11 seconds of padding silence is added before and after the 1 second of spiking. Only this 1 second during which spikes occur is offset---and by no more than the padding silence so that the endpoints of the recording remain unchanged. A sample pair of spike trains is seen on the top right: the original randomly generated train uses the darker line; its copy, offset by 50 ms, appears in the lighter gray. 

On the left, markers indicate the average distance between an unaltered and offset Poisson process across 600,000 realizations of the process. Error bars indicate the sample standard deviation across realizations. While we have constrained parameters and scalings such that the initial rate of increase is consistent across metrics, and so is the maximum value, the plot also shows that the metrics share an extended initial region of linear increase transitioning into a more or less level maximum. Shape differences between the curves exist, particularly between the Victor-Purpura distance and the others, reflecting the sudden transition between the offset-dependent and -independent parts of $D_{V \! P}$'s response. $D^2_{vR}$ and $d_C$ have a more graduated transition and closely agree.

On the bottom right side of Fig.~\ref{Fig:metricDemos}(A), we have the $\sigma_X \,$-convolutions corresponding to the two rasters in the top right (assuming delta function spike area 138 $\mu$V$\cdot \,$s). This illustrates the first implicit step in processing that $d_C$ uses to compare data. Again, the darker line corresponds to the original, and the gray line is the offset version.

In Fig.~\ref{Fig:metricDemos}(B), we see the first of our four plots that use simulated data from a Hodgkin-Huxley neuron as input to $d_C$. Spike train inputs to $D^2_{vR}$ and $D_{V \! P}$ are generated by applying a simple spike detection algorithm to the recorded membrane potential: Peaks in the potential occurring above -20 mV are recorded as spikes.

The input to the neuron is a simple inject current signal, generated by a Gaussian random walk process with a 50 ms time decay toward an equilibrium value of -1.5 $\mu$A/cm$^2$. Steps in the random walk are chosen from a normal distribution with 0 mean and standard deviation $4.0 \, \delta t / \! \sqrt{\delta t / \! \left(50 \mbox{ ms}\right)}$ $\mu$A/cm$^2$, where $\delta t$ is the step size for the simulation (0.01 ms in our case). The power spectrum for fluctuations in this process plateaus below 20 Hz, above which it falls off as $1/ \! \abs{f}^2$. The input offers a simplified representation of the aggregate current a neuron might receive at some point in vivo.

Similarly to the procedure we use for Fig.~\ref{Fig:metricDemos}(A), in the non-offset case the neuron receives this input only during a 1-second on-period that is surrounded on both sides by 0.11 seconds of zero-input padding. The offset input is generated by shifting the on-period only. As the on-period never approaches the end of the padding, the duration of the simulation is fixed. Trials are used if the neuron spikes precisely four times in response to the 1-second on-period so that again we have a neuron spiking at roughly 4 Hz.

Sample membrane potential output for the neuron under a realization of this input is shown in the top right plot of Fig.~\ref{Fig:metricDemos}(B). Once more, the lighter-colored trace is the 50-ms-offset version of the darker trace. The tops of spikes ($\sim \! 40$ mV) are clipped so as to show more subthreshold details in the recording. The convolution of the two sample recordings by $\sigma_X \,$ (for $\tau \! = \! 50$ ms) appears on the bottom right.

Metric distances are plotted on the left versus the input offset. For each offset, the metrics are evaluated on 1,100 independently generated input pairs\footnote{Fewer realizations than in (A) are used because $d_C$ takes a similarly greater time to evaluate on the time-extended membrane potential recordings than any of the metrics do on discrete spike trains.}. Means and sample standard deviations for each metric across realizations are represented in the graph, respectively, by the markers and error bars.

In the plot, we see that, at higher offsets, our convolution distance, $d_C$, indicates a significantly larger distance than do the spike train metrics. Keep in mind that we have normalized these distances to each other, so the excess is not a simple effect of normalization: $d_C$ is genuinely representing about a 100\% higher distance for the larger offsets than are either of the spike train metrics. This is important because spike time differences are only one of the ways that membrane potential recordings differ from each other under this type of input. They also differ in their subthreshold dynamics due to the membrane potential fluctuations produced by the noise current. An agreement between the metrics here would indicate that $d_C$ offers no advantage over $D^2_{vR}$ and $D_{V \! P}$; too much of an excess would indicate an over-emphasis on subthreshold behavior. Instead, we see that $d_C$ monitors both subthreshold and spike time differences, the former of which are ignored by spike time metrics by construction.

In Fig.~\ref{Fig:metricDemos}(C), we reduce the amplitude of the inject current fluctuations by a factor of one half so that random walk steps are chosen from a normal distribution of mean 0 and standard deviation $2.0 \, \delta t / \! \sqrt{\delta t / \! \left(50 \mbox{ ms}\right)}$ $\mu$A/cm$^2$. The input produces substantial subthreshold variations but rarely any spikes. Realizations are chosen such that no spikes at all are produced during the 1-second input period, and the distance is plotted versus the on-period offset. Again, 1,100 realizations are generated for each point on the graph, which represents the mean and sample standard deviation of the distance across realizations. Sample raw and $\sigma_X \,$-convolved membrane potentials are shown on the right. Here we see the response of $d_C$ to purely subthreshold dynamics for the neuron. Naturally, $D^2_{vR}$ and $D_{V \! P}$ have no response as there are no spikes.

In Fig.~\ref{Fig:metricDemos}(D), the same model for the current input as in (C) is used again. This time, however, superthreshold pulse input to the neuron is added. This input takes the form of a sudden increase to the neuron's excitatory synapse conductance\footnote{Increment is  0.1 mS/cm$^2$. 0 mV is used for the excitatory synapse reversal potential.}, which then decays with time constant 5 ms. This is sufficient to produce a spike in almost all cases. Times for these inputs are chosen as a Poisson process at 4 Hz.  Runs are discarded in the event the neuron does not happen to spike in response to all pulses.

In plotting the distances versus the offset, only the current input is offset. The pulse inputs are the same for both the original and offset neural responses. This allows us another way of gauging how significant the subthreshold variations in the membrane potential are to $d_C$: this time in the presence of (fixed) spikes. In the sample membrane potential and convolution plots on the right, we can see that the spikes in the non-offset and 50-ms-offset versions are highly overlapping. On the left, we see that while all metrics respond to current offset, $d_C$'s response is much greater. The spike train metrics respond because the current offset has a small but non-negligible effect on the spike latency in response to the superthreshold pulses. The response of $d_C$ is much greater since the slight spike timing jitter this introduces is a minor effect compared to changes to the membrane potential introduced by the current offset.

Finally, Fig.~\ref{Fig:metricDemos}(E) shows a similar plot to (D), with the difference that instead of having the superthreshold pulse input exactly coincident in both the original and altered versions of the input (the neural response to which the metrics are comparing), we offset the pulses by exactly 5 ms each time. This number does not change throughout the plot. Just as in (D), only the current has its offset varied. The importance is that (E) allows us to see the relative effect of varying the subthreshold current in the presence of fixed non-zero differences in the superthreshold input. As we see, only our convolution distance, $d_C$, shows significant variation over the course of the plot, with the van Rossum and Victor-Purpura distances holding essentially fixed on average. Spike latency effects are not systematically affecting the spike train metrics because while offsets to the current do shift spikes, they make spike offsets smaller just as often as larger. Once more we see that $d_C$ provides access to information that $D^2_{vR}$ and $D_{V \! P}$ do not give us access to. Sample raw and $\sigma_X \,$-convolved membrane potentials appear to the right.

Our demonstrations give us confirmation that, as regards spike-time offsets alone, $d_C$ has a response that is similar to $D^2_{vR}$ and $D_{V \! P}$. We have also seen several situations in which $d_C$ captures a substantial amount of additional, meaningful information regarding differences in the subthreshold behavior of the membrane potential---and therefore differences in the neuron's subthreshold inputs---than what is reported by $D^2_{vR}$ or $D_{V \! P}$.

\section{Discussion}

We have seen that a generalized convolution metric, $d_{gen}$, of the form, Eq.~(\ref{Def:dGen}), has a first order response to both same-time membrane potential differences, $V_1 \! \left(t\right) - V_2 \! \left(t\right)$, and to differences in spike timing between $V_1$ and $V_2$, provided that its kernel, $\sigma$, satisfies the condition, Eq.~(\ref{Cond:SigOK}). After further exploring some of these metrics' response characteristics, both to spikes and to fluctuations of various frequencies in the membrane potential, we obtained a specific choice, $\sigma_X$ (Eq.~(\ref{Def:sigC})), for the kernel that satisfies Eq.~(\ref{Cond:SigOK}), preserves a near-maximal amount of high-frequency information in the membrane potentials, is otherwise free of frequency bias and causes the metric to return 0 only for identical recordings. A kernel that discards high frequency components---more suitable for applications with substantial high frequency noise---and retains the other properties appears in App.~\ref{App:sigNu}.

On the subject of computing $d_C$ for sampled data, one needs to do a quick and accurate job of approximating the integrals involved. The convolution operation, $\int_T\sigma_X \! \left(\left(t-t'\right)/\tau\right) \cdot \left(V_1\! \left(t'\right) - V_2 \! \left(t'\right)\right) dt'/\tau$, requires a separate integral for each point, $t$, within $\tau/2$ of $T$, and is therefore the most intensive part of the computation. Time can be saved by thoughtfully (and sparsely) choosing the points for which the convolution needs to be explicitly evaluated versus the points at which it can be interpolated.

One can also save computational overhead by omitting points from the membrane potential difference which can be linearly interpolated from surrounding points. If one takes this approach, it is crucial to remember that the product, $\sigma_X \! \left(\left(t-t'\right)/\tau\right) \cdot \left(V_1\! \left(t'\right) - V_2 \! \left(t'\right)\right)$, will not be a straight line between the sampled points but rather a more complex (though analytic) function. The second derivative (usually) diverges at the point, $t \! - \! t'=0$, so one should not attempt a trapezoidal integration without sampling the product at this point and in its vicinity. A piecewise analytic integral avoids this necessity. The density of sampled points, and thus the required number of computations, can often be further reduced by using a cubic spline or other polynomial approximation for $V_1 \! \left(t\right)-V_2 \! \left(t\right)$.

It is worthwhile to note that $d_C$ is an inner product metric: it applies the $L^2$ Hilbert space inner product norm to linearly transformed (convolved) membrane potentials. This means that the Hilbert space inner product itself induces a useful quantification:
\begin{align}
& _C  \langle V_1,V_2; \tau\rangle_T \equiv N \! \left[ \sigma_X\right]^2 \int_{-\infty}^{\infty} 
	\widehat{V}_1^{\sigma_X} \! \left(t;\tau\right)_T \cdot
	\widehat{V}_2^{\sigma_X} \! \left(t;\tau\right)_T
	dt/ \! \abs{T}  \notag \\
	& \quad = \frac1{128} \int_{-\infty}^{\infty} \! \left(
	 	\int_T \int_T \sigma_X \! \left(\left(t-t'\right)/\tau\right) V_1 \! \left(t'\right)  \cdot
		  \sigma_X \! \left(\left(t-t''\right)/\tau\right)  V_2 \! \left(t''\right) \frac{dt' dt''}{\tau^2}
	  \right) \! \slfrac{dt}{\abs{T}}
\end{align}
This ``convolution inner product'' has a similar meaning to a dot product between vectors in a finite-dimensional vector space. Since the Cauchy inequality holds for the $L^2$ inner product, we may quantify the ``collinearity'' of two recordings by taking the ratio,
\begin{equation}
\slfrac{_C \langle V_1,V_2; \tau\rangle_T}{\sqrt{_C \langle V_1,V_1; \tau\rangle_T \, \times \,  _C \langle V_2,V_2; \tau\rangle_T}} 
\end{equation}

This ratio has a maximum of unity that occurs strictly for $V_1$ and $V_2$ that differ by at most a multiplicative constant. It decreases with discrepancies in the timing of various features, including spikes, and with local membrane potential displacements up or down. The measure can be made more informative by shifting the membrane potential time courses so that their combined mean is zero before passing them to the inner product. \citet{PaivaEtAl2009} discuss convolution-based inner products for spike trains.

Our metric may be beneficially applied to physiological recordings or computer simulation data any time that spike timing and subthreshold signal differences are both of interest. Some specific applications for $d_C$ include the following:
\begin{enumerate}
\item comparing simulated model neuron behavior to experimentally observed neurons
\item comparing simulated neural network behavior to experimental observations
\item quantifying synchrony among neurons
\item comparing state trajectories of simulated or observed neural systems under differing initial or external conditions
\end{enumerate}

To extend to a metric over time courses for ensembles rather than individual neurons, one can add distances between corresponding neural signals in the ensemble. For example, a root sum of squares (RSS) over distances between the individual signals is appropriate. If it is ambiguous which neurons in the two ensembles correspond to each other, one can apply the metric to the ``bulk,'' ensemble-averaged signals. More refined constructions such as what \citet{HoughtonSen2008} apply to $D_{vR}$ may be applied to $d_C$ as well.

Recent advances in optical methods \citep[e.g.,][]{StPierreEtAl2014, QuirinEtAl2014} and continuing advances in computational neuroscience \citep[see][]{Markram2012} make it possible to obtain simultaneous data for increasingly many neurons \citep{YusteChurch2014}. As this data comes online, it is important to think about ways to process it. The convolution metric, which can be applied to chemical (e.g., calcium) data as well as to membrane potentials, may be particularly relevant for this since it extends so naturally to a metric over neural ensemble data.

By applying the metric to simulated neural data, we have confirmed that it responds in a desirable way to complex membrane potentials, increasing initially linearly with timing offset for several types of randomly generated data, including data containing multiple spikes as well as no spikes. In the process, we have confirmed that $d_C$ provides a considerable amount of information not available from spike train metrics.

The membrane potential contains information about what a neuron is ``hearing'' in addition to what it is ``saying.'' This means that $d_C$ accesses information on differences in the trajectory and computations of the local network for the neurons it is applied to in addition to trajectory and computational differences for the neurons themselves.


\section{Acknowledgements}

I would like to thank John Collins for his steady attention, advice and support throughout the development of the ideas and expression presented here. I am further indebted to Steven Schiff for numerous invaluable suggestions regarding the verbal presentation and this position of this work within the context of important results by others. My gratitude extends also to Reka Albert and Peter Molenaar for their insightful comments and critique and to Jorge Sofo, whose efforts have been crucial to this project's success.


\bibliographystyle{apacite}
\bibliography{ConvolutionMetric}

\newpage
\appendix
\section{Proof of Eq.~(\ref{Eq:InitRiseGGl})}
\label{Appdx:dglRise}

In this appendix, we will prove that:

{\bf Proposition: } \emph{If $\int_{-\infty}^{\infty} \sigma' \! \left(x\right)^2 dx$ converges then for sufficiently small $\abs{x_d}$,}
\begin{align}
\gamma_{gen}\left[\sigma\right]\left(x_d\right)
		\approx N \! \left[\sigma\right] \abs{x_d} \sqrt{\int_{-\infty}^{\infty} \sigma' \! \left(x\right)^2 dx}
\end{align}
\emph{}

{\bf Proof: }

The derivation begins with Eq.~(\ref{Eq:GammaGenSS_sig}):
\begin{align}
	&\gamma_{gen}\left[\sigma\right]\left(x_d\right) = 
		N \! \left[\sigma\right] \sqrt{\int_{-\infty}^{\infty} \left(\sigma \! \left(x\right) 
		- \sigma \! \left(x + x_d\right)\right)^2 dx}	
\end{align}

Differentiating this expression, we find
\begin{align}
	\frac{d}{dx_d}  \gamma_{gen}\left[\sigma\right]\left(x_d\right) 
	& = N \! \left[\sigma\right] 
		\frac{-2\int_{-\infty}^{\infty} \left(\sigma \! \left(x\right) - \sigma \! \left(x + x_d\right)\right)
		\sigma' \! \left(x+x_d\right)dx}{2\sqrt{\int_{-\infty}^{\infty} \left(\sigma \! \left(x\right) 
		- \sigma \! \left(x + x_d\right)\right)^2 dx}} \\
	& = -N \! \left[\sigma\right]^2 
		\frac{\int_{-\infty}^{\infty} \left(\sigma \! \left(x\right) - \sigma \! \left(x + x_d\right)\right)
		\sigma' \! \left(x+x_d\right)dx} {\gamma_{gen}\left[\sigma\right]\left(x_d\right)} 
		\label{Eq:dgdxd}
\end{align}

Consider the one-sided limits,
\begin{align}
	& \lim_{x_d \to 0 \pm} \frac{d}{dx_d}  \gamma_{gen}\left[\sigma\right]\left(x_d\right) \\
	& \qquad = -N \! \left[\sigma\right]^2 \lim_{x_d \to 0 \pm} 
		\frac{\int_{-\infty}^{\infty} \sigma \! \left(x\right) \sigma' \! \left(x+x_d\right) dx
		- \int_{-\infty}^{\infty} \sigma \! \left(x + x_d\right) \sigma' \! \left(x+x_d\right)dx} 
		{\gamma_{gen}\left[\sigma\right]\left(x_d\right)} \\
	& \qquad = -N \! \left[\sigma\right]^2 \lim_{x_d \to 0 \pm} 
		\frac{\int_{-\infty}^{\infty} \sigma \! \left(x-x_d\right) \sigma' \! \left(x\right) dx
		- \int_{-\infty}^{\infty} \sigma \! \left(x\right) \sigma' \! \left(x\right)dx} 
		{\gamma_{gen}\left[\sigma\right]\left(x_d\right)} 
		\label{Eq:limdgdxd}
\end{align}

Assuming $\sigma$ is continuous, both numerator and denominator of Eq.~(\ref{Eq:limdgdxd}) go to zero in these limits, in which case we may apply L'H\^opital's rule:

\begin{align}
	& \begin{aligned}
		\lim_{x_d \to 0 \pm} \frac{d}{dx_d}  \gamma_{gen}\left[\sigma\right]\left(x_d\right) = 
		& - N \! \left[\sigma\right]^2 \left( \lim_{x_d \to 0 \pm}
		\frac{d}{dx_d}  \gamma_{gen} \left[\sigma\right] \left(x_d\right)\right)^{-1} \\
		& \cdot \lim_{x_d \to 0 \pm} \frac{d}{dx_d} \int_{-\infty}^{\infty} 
		\left(\sigma \! \left(x\right) - \sigma \! \left(x + x_d\right)\right) \sigma' \! \left(x+x_d\right)dx	
	\end{aligned} \\
	& 	\begin{aligned}
			\Rightarrow \left(\tfrac1{N \! \left[\sigma\right]} \lim_{x_d \to 0 \pm} \frac{d}{dx_d}  
			\gamma_{gen}\left[\sigma\right]\left(x_d\right)\right)^2	  
			= &-\lim_{x_d \to 0 \pm} \frac{d}{dx_d} \int_{-\infty}^{\infty} 
			\sigma \! \left(x\right) \sigma' \! \left(x+x_d\right)dx \\
			&+ \lim_{x_d \to 0 \pm} \frac{d}{dx_d} \int_{-\infty}^{\infty}
			\sigma \! \left(x + x_d\right) \sigma' \! \left(x+x_d\right) dx
		\end{aligned} \\
	&\qquad = -\lim_{x_d \to 0 \pm} \left[\frac{d}{dx_d} \int_{-\infty}^{\infty} 
			\sigma \! \left(x-x_d\right) \sigma' \! \left(x\right)dx 
		 	+ \frac{d}{dx_d} \int_{-\infty}^{\infty}
			\sigma \! \left(x\right) \sigma' \! \left(x\right) dx \right] \\
	&\qquad = \lim_{x_d \to 0 \pm} \int_{-\infty}^{\infty} \sigma' \! \left(x - x_d\right)\sigma' \! \left(x\right) dx +0\\
	&\qquad = \int_{-\infty}^{\infty} \sigma' \! \left(x\right)^2 dx \\
	& \Rightarrow \lim_{x_d \to 0 \pm} \frac{d}{dx_d}  \gamma_{gen}\left[\sigma\right]\left(x_d\right)
		= \pm  N \! \left[\sigma\right] \sqrt{\int_{-\infty}^{\infty} \sigma' \! \left(x\right)^2 dx} \\
	& \Rightarrow \gamma_{gen}\left[\sigma\right]\left(x_d\right)
		\approx N \! \left[\sigma\right] \abs{x_d} \sqrt{\int_{-\infty}^{\infty} \sigma' \! \left(x\right)^2 dx} 
		\ : \ \abs{x_d} \mbox{ sufficiently small}
\end{align}
The last two conclusions follow since $\gamma_{gen}$ has a minimum of 0 at $x_d=0$.

This proof assumes continuity for $\sigma$ and that integrals of the form $\int_{-\infty}^{\infty} \sigma \! \left(x - x_d\right) \sigma' \! \left(x\right)dx$ are differentiable. The convergence of $\int_{-\infty}^{\infty} \sigma' \! \left(x\right)^2 dx$ entails both since squares of Dirac delta functions cannot be integrated and since:
\begin{equation}
\abs{\frac{d}{dx_d} \int_{-\infty}^{\infty} \sigma \! \left(x - x_d\right) \sigma' \! \left(x\right)dx} = \abs{\int_{-\infty}^{\infty} \sigma' \! \left(x - x_d\right) \sigma' \! \left(x\right)dx} \le \abs{\int_{-\infty}^{\infty} \sigma' \! \left(x\right)^2 dx}
\end{equation}


\section{Triangle Inequality for $d_{gen}$}
\label{Sec:TriangleInequality}

The triangle inequality for $d_{gen}$ will be more accessible if we first prove that:

\emph{For any three bounded functions, $x \! \left(t\right), y \! \left(t\right), z \! \left(t\right)$ with $\abs{z \! \left(t\right)} \le \abs{x \! \left(t\right) + y \! \left(t\right)}$:}
\begin{equation}
\sqrt{\int_T x \! \left(t\right)^2dt} +\sqrt{\int_T y \! \left(t\right)^2dt} \ge \sqrt{\int_T z \! \left(t\right)^2dt} 
\label{LemmaTrIn}
\end{equation}
\emph{where $T$ is a finite domain.}

This inequality follows easily from the Cauchy inequality for $L^2$ Hilbert spaces:
 \begin{align}
\int_{-\infty}^\infty x \! \left(t\right)^2dt\int_{-\infty}^\infty y \! \left(t'\right)^2dt' \ge \left(\int_{-\infty}^\infty x \! \left(t\right)y \! \left(t\right)dt\right)^2
\end{align}
for square-integrable $x \! \left(t\right)$ and $y \! \left(t\right)$.

Taking the square root, we have, for any bounded functions, $x \! \left(t\right)$, $y \! \left(t\right)$,
\begin{align}
\sqrt{\int_T x \! \left(t\right)^2dt\int_T y \! \left(t'\right)^2dt'} \ge \int_T x \! \left(t\right)y \! \left(t\right)dt 
\end{align}

Doubling both sides and adding $\int_T x \! \left(t\right)^2 dt + \int_T y \! \left(t\right)^2 dt$ gives:
\begin{align}
\int_T x \! \left(t\right)^2 dt + \int_T y \! \left(t\right)^2 dt + 2\sqrt{\int_T x \! \left(t\right)^2dt\int_T y \! \left(t'\right)^2dt'} \ge \qquad & \notag \\
 \int_T x \! \left(t\right)^2 dt + \int_T y \! \left(t\right)^2 dt + 2\int_T x \! \left(t\right)y \! \left(t\right)dt & 
\end{align}
\begin{align}
&\Rightarrow \left(\sqrt{\int_T x \! \left(t\right)^2 dt} + \sqrt{\int_T y \! \left(t\right)^2 dt} \right)^2 \ge \int_T\left(x \! \left(t\right) + y \! \left(t\right)\right)^2 dt \\
&\Rightarrow \sqrt{\int_T x \! \left(t\right)^2 dt} + \sqrt{\int_T y \! \left(t\right)^2 dt} \ge \sqrt{\int_T z \! \left(t\right)^2 dt}
\end{align}

With this inequality accessible, it is straightforward to prove:

{\bf Proposition, Triangle inequality for $d_{gen}$:}

\emph{Given three well-behaved time courses, $V_1 \! \left(t\right)$, $V_2 \! \left(t\right)$, $V_3 \! \left(t\right)$, and a bounded kernel, $\sigma \! \left(x\right)$:}
\begin{equation}
d_{gen}\left[\sigma\right]\left[V_1 \! \left(t\right), V_2 \! \left(t\right);\tau\right]_T
	+ d_{gen}\left[\sigma\right]\left[V_2 \! \left(t\right), V_3 \! \left(t\right);\tau\right]_T
	\ge d_{gen}\left[\sigma\right]\left[V_1 \! \left(t\right), V_3 \! \left(t\right);\tau\right]_T
\end{equation}

{\bf Proof: } Define the functions, $s_{12} \! \left[\sigma\right]_T \! (t)$, $s_{23} \! \left[\sigma\right] \! (t)_T$, and $s_{13} \! \left[\sigma\right] \! (t)_T$, according to:
\begin{align}
s_{ij}\left[\sigma\right] \! \left(t\right)_T
	&\equiv \int_T \sigma \! \left(\left(t - t'\right)/\tau\right)
	\cdot \left(V_i \! \left(t'\right) - V_j \! \left(t'\right)\right) dt'/\tau
\end{align}
This gives:
\begin{equation}
d_{gen}\left[\sigma\right]\left[V_i \! \left(t\right), V_j \! \left(t\right);\tau\right]_T = N\left[\sigma\right] \sqrt{\int_{-\infty}^{\infty} s_{ij} \! \left[\sigma\right] \! \left(t\right)_T^{\, 2} dt/ \! \abs{T} }
\end{equation}

We will assume that well-behaved $V_i$ means the $s_{ij}$ are bounded. Recognizing that:
\begin{align}
s_{13} \! \left(t\right)_T & = s_{12} \! \left(t\right)_T + s_{23} \! \left(t\right)_T \\
\Rightarrow \abs{s_{13} \! \left(t\right)_T} & \le \abs{s_{12} \! \left(t\right)_T + s_{23} \! \left(t\right)_T}
\end{align}
the proposition follows from Eq.~(\ref{LemmaTrIn}).

\section{Kernel for Noisy Data}
\label{App:sigNu}

In the presence of significant detection noise, i.e., noise that is not intrinsic to the neuron or network of neurons being observed but rather originates in the act of observation \emph{per~se}, one may prefer a metric that disregards high frequency information rather than preserves as much of it as possible. Nonetheless, one still wants the metric to avoid bias toward or against specific frequency bands and to yield zero strictly for identical inputs (identity of indiscernibles). The following kernel, $\sigma_N$, is applicable to such a case:
\begin{equation}
\sigma_N \! \left(x\right) \equiv
	\begin{cases}
		\frac{36}{10} - 192 \, x^2 + 768 \left( \abs{x}^3 - x^4 \right) - \frac{3072}{10} \, \abs{x}^5
			&: \abs{x} \le \frac14 \\
		\frac{48}5 \left(1 - 2 \abs{x} \right)^5 &: \frac14 < \abs{x} \le \frac12 \\
		0 &: \ \mbox{otherwise}
	\end{cases}
\end{equation}

This piecewise polynomial satisfies our main restriction for kernels, Eq.~(\ref{Cond:SigOK}), with $\int_{-\infty}^{\infty} \sigma'_N \! \left(x\right)^2 dx = \frac{3512}{35}$, giving $N \! \left[\sigma_N\right] = \frac12 \sqrt{\frac{35}{878}}$. The Fourier transform for $\sigma_N$ is:
\begin{align}
& \tilde{\sigma}_N \! \left(k\right) = \frac{\ 36}{\sqrt{2 \pi}} \ 
	\frac{\left( 1 - \sinc\left(k/4\right) \right)^2}{(k/4)^4}
	\mbox{,} \quad \mbox{where $\sinc{x} \equiv \slfrac{\sin{x}\,}{\,x} $}
\end{align}
This $\tilde{\sigma}_N$ is zero-free, oscillation-free and rolls off as $1/k^4$ owing to $\sigma_N$'s continuous second derivative. This is much faster than the other kernels we have discussed.

The kernel is the result of a two-step construction process. First a cusped kernel, $\sigma_M$, obtains by integrating a triangular kernel across scales, weighting by the scale. As with $\sigma_X$, this makes for a kernel free of local minima and zeroes in its Fourier transform:
\begin{align}
\sigma_M \! \left( x \right) & \equiv
	\begin{cases}
		3 \int_{2\abs{x}}^1 x_0 \, 2 \left(1 - 2 \abs{x}/x_0\right) dx_0 
		= 3 \left(1 - 2\abs{x}\right)^2 &: \abs{x} \le \frac12 \\
		0 &: \ \mbox{otherwise}
	\end{cases} 
\end{align}

The smooth $\sigma_N$ kernel is a rescaled autoconvolution of $\sigma_M$:
\begin{equation}
\sigma_N \! \left(x\right) =
	4 \int_{-\infty}^{\infty} \sigma_M \! \left( 2x' \right) \sigma_M \! \left( 2 \left(x - x'\right) \right) dx'
\end{equation}

\section{Estimating the spike area $\alpha$ for the simulated Hodgkin-Huxley neuron}
\label{App:alphaEstimate}
In Sec.~\ref{Sec:Demonstrations}, it is necessary to estimate the effective area of the simulated Hodgkin-Huxley neuron's spikes to produce Fig.~\ref{Fig:metricDemos}. I make this estimate by repeatedly evaluating our convolution metric, $d_C \! \left[V_1,V_2;\tau\right]$, on pairs of simulated membrane potentials, both of which contain a single spike, and varying the time difference, $t_d$, between the spikes. The time sensitivity, $\tau$, is set to $\tau_X=50$ ms, the same value we use for $d_C$ in Sec.~\ref{Sec:Demonstrations}. Performing a linear regression on $d_C$ versus $\gamma_X \! \left(t_d/\tau_X\right)$ produces the estimate for $\alpha$.

A separate estimate for $\alpha$ is made for each of the panels, (B--E), in Fig.~\ref{Fig:metricDemos}: the differences from panel to panel in the type and nature of the neuron's input could have some effect on the average area of its spikes. For each estimate, I generate sections of noise current input 0.25 seconds in length according to the same process that generates the noise current applied to the neuron in the relevant panel. These are down-selected on the requirement that the neuron does not spike during the 0.25 seconds. A superthreshold synaptic pulse is then added to the middle of the input, producing a single spike in the simulated neuron there. A second recording is produced according to the same procedure with the pulse shifted by an amount between 0--25 ms. Spike times are extracted from both recordings via a simple spike detection algorithm which labels local maxima above -20 mV as spikes. 2,200 pairs of recordings are produced in this way, not all of which are used: sometimes one or both recordings contain a number of spikes other than one, in which case the pair is discarded.

For the non-discarded recording pairs, I record the difference, $t_d$, between the spike times, and evaluate the convolution distance, $d_C \! \left[V_1,V_2;\tau_X\right]$, between the recordings. With these values in hand, a collection of ordered pairs, $\left(\gamma_X \! \left(t_d\right)\!,\, d_C\right)$ is then constructed. Regressing the distance, $d_C$, versus $\gamma_X \! \left(t_d\right)$, produces a regression coefficient that, when multiplied by $\sqrt{\tau_X \abs{T}}$, where $\abs{T} \! = \! 0.25$ s, gives an estimate of $\alpha$ for the simulated neuron valid for the form of input used. This estimate is then used to perform the scalings of $D^2_{vR}$ and $D_{V \! P}$ for that type of input (corresponding to one of the panels in Fig.~\ref{Fig:metricDemos}). For each input type, the estimate is close to 138 $\mu$V$\cdot \,$s, with estimation error $\sim$1\%.

The contribution of the estimation error for $\alpha$ to the error on the scaled versions of the spike train metrics is included in the spike train metrics' error bars in Fig.~\ref{Fig:metricDemos}.

\end{document}